Preprinted at arXiv:2503.22916 (https://arxiv.org/abs/2503.22916)

After modified this preprint, accepted (18 August 2025) at ***The Innovation Life***
https://doi.org/10.59717/j.xinn-life.2025.100155 (please check the final accepted paper)

Preprinted manuscript

# Anti-pathogenic property of thermophile-fermented compost as a feed additive and its *in vivo* external diagnostic imaging in a fish model

Hirokuni Miyamoto[1,2,3,4,5*], Shunsuke Ito[6], Kenta Suzuki[7,8], Singo Tamachi[9], Shion Yamada[10], Takayuki Nagatsuka[3,4], Takashi Satoh[11], Motoaki Udagawa[12], Hisashi Miyamoto[13], Hiroshi Ohno[2], and Jun Kikuchi[5,14*]

*Affiliations:*
1. *Graduate School of Horticulture, Chiba University, Matsudo, Chiba 271-8501, Japan*
2. *RIKEN Center for Integrative Medical Sciences, Yokohama, Kanagawa 230-0045, Japan*
3. *Japan Eco-science (Nikkan Kagaku) Co. Ltd., Chiba, Chiba 263-8522, Japan*
4. *Sermas Co., Ltd., Ichikawa, Chiba 272-0033, Japan*
5. *Graduate School of Medical Life Science, Yokohama City University, Yokohama, Kanagawa 230-0045, Japan*
6. *Chuubushiryo Co. Ltd., Oobu, Aichi 474-0011, Japan*
7. *RIKEN BioResource Research Center, Tsukuba, Ibaraki 305-0074, Japan*
8. *Institute for Multidisciplinary Sciences, Yokohama National University, Yokohama, Kanagawa 240-8501, Japan*
9. *Center for Frontier Medical Engineering, Chiba University, Chiba, Chiba 263-8522, Japan*
10. *Faculty of Engineering, Chiba University, Chiba, Chiba 263-8522, Japan*
11. *Division of Hematology, Kitasato University School of Allied Health Sciences, Sagamihara, Kanagawa 252-0373, Japan*
12. *Keiyo Gas Energy Solution Co. Ltd., Ichikawa, Chiba 272-0033, Japan*
13. *Miroku Co. Ltd., Kitsuki, Oita 873-0021, Japan*
14. *RIKEN Center for Sustainable Resource Science, Yokohama, Kanagawa 230-0045, Japan*

* Co-corresponding authors
Correspondence: Hirokuni Miyamoto Ph.D., Chiba University, RIKEN, Sermas Co., Ltd., and Japan Eco-science Co., Ltd.
Tel: +81-43-290-3947, Fax: +81-43-290-3947
E-mail: hirokuni.miyamoto@riken.jp, h-miyamoto@faculty.chiba-u.jp

Correspondence: Jun Kikuchi Ph.D., RIKEN
Tel: +81-43-290-3942, Fax: +81-43-290-3942
E-mail: jun.kikuchi@riken.jp

## Abstract
Fermentative recycling of organic matter is important for a sustainable society, but the functionality of fermented products needs to be adequately evaluated. Here, we clarify the antipathogenic properties for fish of a compost-type feed additive fermented by thermophilic *Bacillaceae* using non-edible marine resources as raw materials. After prior administration of the compost extract to seabream as a fish model for 70 days, the mortality rate after 28 days of exposure to the fish pathogen *Edwardsiella* reached a maximum of 20%, although the rate was 60% without prior administration. Under such conditions, the serum complement activity of seabream increased, and the recovery time after anesthesia treatment was also fasten. Furthermore, the differences in the degree of smoothness and glossiness of the fish body surface depending on the administration were statistically shown by imaging techniques to evaluate the texture and color tone of field photographs. These results suggest that thermophile-fermented compost is effective as a functional feed additive against fish disease infection, and that such conditions can be estimated by body surface analysis. This study provides a new perspective for the natural symbiosis industry, as well as for the utilization of non-invasive diagnosis to efficiently estimate the quality of its production activities

*Abbreviations*: GLCM, Gray Level Co-occurrence Matrix for texture analysis; HSV, Heu (color tone), Saturation (colorfulness) and Value (brightness)

## Introduction
Planetary boundaries[1], a concept that states the limits of human survival on Earth, were defined in 2009, and thereafter, the severity of the limits has increased. The excessive influx of nitrogen, phosphorus, new chemicals, and artificially synthesized antimicrobials into the biosphere, followed by loss of biodiversity, has already reached dangerous levels[2]. In the light of these issues, a global goal should be set to move towards a recovery trend from biodiversity loss, that is, "nature positive" [3].
To address these issues, it is necessary to adopt a comprehensive approach that integrates science and technology. For example, organic matter recycling technology is important for the effective use of nitrogen and phosphorus[4] as environmental science technology. However, to promote the use of compost, it is necessary to evaluate its functionality. The most popular examples of the effective use of compost made from recycled organic matter are agricultural use[5-7], and there are also cases of use as feed[8-10]. The main theme of these applications is from the viewpoint of efficient resource recycling of fermented products. Among new chemical substances, antibiotics can lead to the spread of resistant bacteria; therefore, alternative technologies are needed, along with appropriate restrictions on their use. Furthermore, even if new alternatives are found from these technical perspectives, it will take time to disseminate them because the evaluation of functionality for animals and plants requires detailed analysis, which is generally expensive and time-consuming. Therefore, a simple combined evaluation technique[11] is necessary to contribute to a sustainable society at an early stage.
In recent years, the multiphasic function as an organic fertilizer

[12-14] or feed additive[15-19] of fermented compost produced from non-edible marine animal resources with thermophilic *Bacillaceae* using closed fed-batched bioreactors under high-temperature conditions (70-80 °C) has been investigated (hereinafter, thermophile-fermented compost, but not fecal fermented compost). As a feed additive, compost has been shown to alter the gut microbiota of insects [20,21], rodents, fish, and livestock animals, followed by the improvement of the productivity of chickens and pigs, reduction of stillbirth rates in pigs, and improvement of the immune system [18] and antioxidant activity [19] in rodents as an animal model. In fish, improvement in muscle quality (increase in free amino acids) of carp[16] and flounder [15] have been observed, and seagrass overgrowth has been confirmed in and around the drainage channels of the aquaculture facilities for flounder [22]. In these facilities, the mortality rate has also improved.

In this study, we assessed the effects of thermophile-fermented compost on infectious diseases using seabream as a fish model. The assessment proceeded with an evaluation inside and outside the body (*i.e.*, an *in vivo* internal and external evaluation). In particular, since it has long been known that the surface of the fish body changes depending on environmental conditions[23-25], the surface was analyzed by photographs via newly created commands using diagnostic imaging techniques[26-33], which have been attracting attention in recent years. Thus, the *in vivo* external evaluation of the fish body surface could estimate the physiological responses based on the *in vivo* internal evaluation, and the utilization of the thermophile-fermented compost for fish infectious diseases was shown. This study suggests the existence of a compost that can efficiently recycle nitrogen and phosphorus and also help to reduce artificial synthetic antimicrobials, and provides the possibility of efficiently evaluating these multiple effects with diagnostic imaging using field photographs.

## Materials and Methods

### *Fish breeding conditions*

Fish tests were conducted using seabream at the Oigawa proving grounds in Japan (N34.80, E138.31) (Chubushiryo Co., Ltd., Japan) between February and June. Experiments were conducted according to the animal care guidelines of the farm. In brief, in a 0.5 t capacity of a round tank circulating in one direction with seawater, the temperature of the round tank water was maintained at approximately 22 °C. The seabreams (at most 15 per tank) were fed commercially available extruded pellet (EP) feed (Tai Next EP; Chubushiryo Co., Ltd., Japan) with 0% (group name: Gr1), 1% (group name: Gr2), and 5% (group name: Gr3) (weight/weight) of the water solution of thermophile-fermented compost (compost extract). The thermophile-fermented compost was prepared as follows. Fermented compost from marine animal resources (MAR compost) was produced by an aerobic repeated fed-batch fermentation system at high temperatures (approximately 70-80°C) via fermentation-associated self-heating as previously described[12] (Miroku Co., Ltd. and Keiyo Gas Energy Solution Co., Ltd., Japan). Powdered compost was diluted to 1/100 with water (vol. /vol.) and incubated under aerobic conditions at 60°C for at least 10 h (this solution was used as the compost extract)[15-19,34]. A diluted compost solution was used as the compost extract in this study.

### *In vivo internal detection*

Blood urea nitrogen (BUN), glucose (GLU), triacylglyceride (TG), total cholesterol (T-chol), and total protein (T-Pro) levels in blood samples were determined using an automated biochemical analyzer SPOTCHEM™EZ SP-4430 (ARKRAY Co., Ltd., Japan). The free amino acid content was analyzed in the muscle of the occipital part of the fish obtained after the test. These muscle samples were homogenized with phosphate buffer, and the supernatant (1.5 ml) was poured into a Microcon 3000 column (Millipore Co., Ltd., USA). The filtrates were analyzed using a capillary electrophoresis system (CE system G1600A) equipped with a fused silica capillary with a basic anion buffer, as previously reported. Serum complement activity was determined by assessing the hemolysis rate ACH50, as alternative complement pathway hemolytic activity[35]. Recovery tests were performed in seawater supplemented with 2-methylquinoline (Wako Pure Chemical Co., Ltd., Japan) at the manufacturer's recommended amount per body weight, and the time to wake up from anesthesia was measured to assess the health of the fish body.

### *Artificial infection test*

The seabreams were fed 0% (Gr1), 1% (Gr2), and 5% (Gr3) (weight/weight) of the water solution of compost extract for 70 days, followed by artificial infection of the fish (n=10 per group) intraperitoneally (*i.p.*) with 0.3 mL (conc. $4.3 \times 10^7$ CFU/mL) of *Edwadsiella tarda* (E05-92; Chubushiryo Co., Ltd., Japan) was used for the survival tests. Subsequent breeding was performed for 28 days. The obtained survival data were visualized as survival curve via the function 'ggsurvplot' of the R software libraries "survival" and "survminer" using the Kaplan-Meier method. These difference between the groups were evaluated by the logrank test and Likelihood ratio test via the function 'coxph' of the library "survival". These data were represented via the library "broom", and forest plot based on the cox regression was visualized via the function 'ggforest'.

### *Photographed conditions*

Individuals different from the infection test (Gr1, n=10; Gr2, n=10; and Gr3, n=8) were filmed in May before the end of the test. Photographs were taken immediately after sacification for filming. The film conditions were set as follows: Filming equipment, Canon IXY DIGITAL 50; Color profile, sRGB IEC61966-2.1; size, 1600×1200; focus distance, 5.8 mm; alpha channel, no; metering mode, pattern; F value, f/2.8; and exposure time, 1/50. The original and modified photographs were stored in the figshare site (10.6084/m9.figshare.28646039), and the other information has also been on record. The color change of the background and the dotting of the photographs for the *in vivo* external evaluation (body length evaluation, texture analysis, and HSV analysis) were performed using the Adobe Express (https://www.adobe.com/jp/express/feature/image/remove-background) and the Adobe Photoshop 2025 (version 26.4.1). Gaussian treated images were modified based on the commands described in the GitHub site (https://github.com/kecosz/cv2_jpg_preprocessing) and DOI

(10.5281/zenodo.15081995).

### Body length evaluation

Open CV was performed using Python (version 3.10.8) on Mac OS Sequoia (version 15.3) on arm64. The libraries and modules such as "os", "cv2", "numpy", "math", and "pandas", were imported, and the body length ratio was obtained according to the original commands. The procedure is shown in Figs. 3a and 3b, and the GitHub site (https://github.com/hmiyamoto2000/program_tai3-main/tree/v0.1.1; https://github.com/hmiyamoto2000/program_tai1_com) and DOI (10.5281/zenodo.15080985).

### Body surface texture analysis

Gray Level Co-occurrence Matrix (GLCM) as texture analysis has developed mainly in the medical engineering for the evaluation of image surfaces[26-33,36]. Here, on referring the following the URL (https://github.com/AIM-Harvard/pyradiomics/tree/master/radiomics), new developed commands were performed using Python (version 3.7.16) on Mac OS Ventura (version 13.6.7) on Intel (x86_64), and Python (version 3.10.8) on Mac OS Sequoia (version 15.3) on arm64. The libraries and modules such as "os", "cv2", "numpy", "math", "pandas", "affine", "tifffile", "csv", "itertools", "SimpleITK", "scipy", "Image", "ImageDraw", "tqdm", "glszm", "feature", "my_makedirs", "calc_glcm_mean", "make_img_mask", "calculation", and "glszm_t" were imported. The procedure is shown in Figs. 4a and 4b and at the GitHub site (https://github.com/hmiyamoto2000/program_tai1_Texture_main/tree/v1.0.6) and DOI (10.5281/zenodo.15081663).

### Feature selection

To extract feature texture factors using machine learning algorithms, Random Forest [37], one of machine learning with bagging (bootstrap aggregating), and XGBoost [38], machine learning with Extreme Gradient Boosting [11], and lightgbm for a highly efficient gradient boosting decision tree[39], were performed. These machine learning algorithms can be used not only for predicting test data based on training data, but also for selecting features to classify between groups (feature selection) [11,40,41]. The following libraries and modules were used via Python (version 3.10.8) on Mac OS Sequoia (version 15.3) on arm64: "xgboost", "RandomForestClassifier", "lightgbm", "train_test_split", "accuracy_score", "pyplot", "pandas", "yaml", and "csv". These selected feature factors were visualized as a bubble chart using the "matplotlib" (https://matplotlib.org/stable/index.html).

### HSV analysis

Open CV was performed using Python (version 3.7.16) on Mac OS Ventura (version 13.6.7) on Intel (x86_64), and Python (version 3.10.8) on Mac OS Sequoia (version 15.3) on arm64. The libraries and modules such as "cv2", "numpy", "matplotlib", "tqdm", "scipy.stats", "pandas", "scikit_posthocs", and "pairwise_turkeyhsd" were imported. The procedure is shown in Figs. 6a and 6b for the GitHub site (https://github.com/hmiyamoto2000/program_tai1_HSV_main/tree/v1.0.5) and DOIs (10.5281/zenodo.15081661).

### Statistical analyses

The following statistics were performed for frequentist equivalence testing: $p$ values via the Shapiro-Wilk test, F test, and Bartlett test were used to select parametric and non-parametric analyses. In addition, ANOVA followed by Tukey's HSD and Kruskal-Wallis one-way analysis of variance followed by the Steel-Dwass test were performed as appropriate methods depending on the data sets. For the comparison between Gr1 and a combined compost extract treated group (Gr2 and Gr3), the unpaired t test or the Welch's t test was used for parametric analysis, or the Wilcoxon rank-sum test for nonparametric analysis. Significance was declared when $P < 0.05$, and tendency was assumed at $0.05 \leq P < 0.20$. These calculation data were prepared by using the Python (version 3.10.8) and R software (version 4.4.2). Prism software (version 10.3.1) was used for graphical visualization. Data are presented as mean ± SE. Data collection and analysis were not blinded to experimental conditions.

## Results

### Fish assessment overview

In this study, the effects of administering the diluted extract of thermophile-fermented compost (compost extract) to seabream were examined (Fig. 1a). As shown in Fig. 1b, the schedule consisted of 70 days of pretreatment followed by injection of the pathogen *Edwardsiella tarda*; and subsequently, the survival test was performed for 28 days. In addition, the physiological response and body surface were also examined (an *in vivo* internal evaluation and field photo diagnostic imaging as an *in vivo* external evaluation in Fig. 1c). Notably, it has been observed that the administration of compost extract improved the surface gloss of the fish (Gr2 and Gr3 rather than Gr1) (Fig. 1d). Although no significant differences were observed in these breeding outcomes (Figs.1e and S1), a reduction in the number of deaths was confirmed in the survival test.

### Evaluation of physiological responses in a fish model

The Kaplan-Meier survival curve indicated a significant difference among the three groups (logrank test, $p$=0.02; likelihood ratio test, $p$=0.03). In particular, a significant difference ($p$=0.049) was observed between Gr1 (0% compost extract) and Gr2 (1% compost extract), and the survival rate of Gr2 was higher than that of Gr1. The survival rate of Gr3 (5% compost extract) was also higher than that of Gr1 ($p$ =0.11). The forest plot based on the cox regression (Fig. S2a) also showed the difference as follows: intervention Gr2, $p$=0.0453; and intervention Gr3, $p$=0.0931. The observations indicated that the survival data were independent of the dose of compost extract. The average body weights of all groups decreased after the test, but no significant differences were observed (Fig. S2b). The fork length and the condition factor did not change. In the serum and muscle immediately before the survival test (0d), the levels of serum BUN, Glu, TG, T-Chol, and T-Pro were not significantly different among the three groups. The levels of muscle-free acids (glutamate and glycine) and lactate were also not significantly different. Differences between Gr1 and G2_G3 were significantly observed in complement activity, which is one of the immune indicators ($p$=0.0255), the activity levels of Gr2_Gr3 were raised. Such a tendency was observed between the individual groups (G1 and Gr2; and Gr1 and Gr3), although the difference was not always observed ($p$=0.120 and $p$=0.145). However, these trends were not always maintained until the end of the trial. Furthermore, as a result of verifying the effect of waking up from anesthesia, although not statistically significant ($p$=0.097 and $p$=0.212), it was confirmed that the compost extract groups (Gr2 and Gr3) could significantly wake up quickly from anesthesia ($p$=0.0306), although no significant differences were observed between the individual groups (G1 and Gr2, p=0.097; and Gr1 and Gr3, $p$=0.212). Thus, physiological changes in the effects of compost extract administration were confirmed by *in vivo* internal evaluation, although not always dose dependent.

### *Evaluation of fish body length*

Assessing the dimensions of the fish body is important for understanding the condition of the fish, but it is difficult to perform accurate quantification in field operations. However, it is relatively easy to take photographs and evaluate them at a later date. Here, a command was developed to assess by calculating the ratio for each part when there were no standard dimensions, such as a yardstick, and the operation was performed. As shown in Figs.3a and 3b, it is possible to calculate the ratio from the mouth to the caudal fin (X1-X2) after setting a reference line and the related points in different RGB colors. The ratio of the orthogonal line of the pectoral fin to X1-X2 and the ratio of the orthogonal line of the anus were calculated. These commands were deposited at the GitHub site as described in the Data Availability section. The results confirmed that the target fish body significantly differed in the ratio of R1 to R2 (Fig. 3c). That is, it was confirmed that Gr2, which had the lowest mortality rate, increased the ratio of R1 ($p$=0.0418925) and R2 ($p$=0.0503825). The ratios of Gr3 were significantly different from those of Gr2 (R1, $p$=0.0247454; R2, $p$=0.0376840). No statistically significant differences were observed for R3.

### *Texture evaluation of fish body surface*

As shown in Fig. 1d, the fish body surface appeared to differ depending on the test group. However, a quantitative evaluation method for such differences was not established in the fishery industry, and it is difficult to perform accurate quantification in field operations, even if there is some way. Therefore, the surface of the fish body was evaluated using texture analysis technology[26-33] mainly used in the medical engineering. As shown in Figs.4a and 4b, to evaluate the fish according to the same criteria, we developed a command to set reference points at the tip of the fish's mouth and caudal fin and adjust the points so that they are horizontal. In addition, the background-modified images were also prepared. Next, 1/3 of the central part of the fish body was automatically evaluated as the site to be measured via the new developed commands. These commands were deposited at the GitHub sites and the DOIs as described in the Data Availability section. In addition, because the photographs of the site are not always beautiful, the image data were edited using the Gaussian model, and an evaluation in the case of blurry images was also carried out. Three machine learning algorithms were used to select the feature metrics for the texture analysis data. The algorithm here is used not to predict the accuracy of the test data based on the training data, but to improve the selection accuracy of the feature metrics. By analyzing the original photographs, lightGBM selected eight feature metrics (Fig. S4). The highest features were shown in Fig.4c. Among these metric groups, the metric with higher values for both Gr2 and Gr3 than for Gr1 was GLSZM_ZoneEntropy (Gr1 vs. Gr2, $p$=0.0124425; Gr1 vs. Gr3, $p$=0.0480842; and Gr2 vs. Gr3, $p$=0.0030646). On the other hand, GLSZM_SmallAreaEmphasis (Gr1 vs Gr2, $p$=0.0435837; Gr1 vs Gr3, $p$=0.0000187; and Gr2 vs Gr3, $p$=0.0083970), GLSZM_SizeZoneNonUniformityNormalized (Gr1 vs Gr2, $p$=0.0249181; Gr1 vs Gr3, $p$=0.0000048; and Gr2 vs Gr3, $p$=0.0040308), and LBP_R1_P8_ave (Gr1 vs Gr2, $p$=0.0754322; and Gr1 vs Gr3, $p$=0.0325014) were metrics that led to lower values (Figs. 5a and S5a). In addition, the other metrics did not show statistically significant differences. Next, as a result of the analysis based on the Gaussian image, GLSZM_ZoneEntropy, GLSZM_SmallAreaEmphasis, and LBP_R1_P8_ave were detected among these important metrics, and the trend was similar as that in the original image (Fig. 5b). GLSZM_ZoneEntropy for Gr2 and Gr3 was significantly higher than that for Gr1 (Gr1 vs. Gr2, $p$=0.0015341; Gr1 vs. Gr3, $p$=0.0000018), and GLSZM_SmallAreaEmphasis for Gr2 and Gr3 was significantly lower than that for Gr1 (Gr1 vs. Gr2, $p$=0.0188099; and Gr1 vs. Gr3, $p$=0.0053235) (Fig. 5b), and LBP_R1_P8_ave showed statistical significance in some areas (Gr1 vs. Gr2, $p$=0.0235563; and Gr1 vs. Gr3, $p$=0.2313297) (Fig. S5b). In the Gaussian image, GLSZM_LargeAreaLowGrayLevelEmphasis was a newly important metric candidate, but it was not statistically significant (Fig. S5b). There were no statistically significant differences in the other factors.

Thus, GLSZM_ZoneEntropy should be an essential metric under these experimental conditions.

### *HSV evaluation of fish body surface*

In on-site photography, it is difficult to adjust the light. However, regardless of the degree of light, differences can be recognized by human vision, and these differences may be visible to those who have abundant experience in the field. Assessing the difference is important for understanding the condition of fish, but it is difficult to perform accurate quantification in field operations. Therefore, the surface of the fish body was evaluated via HSV analysis. As shown in Figs.5a and 5b, a survey was conducted on the part analyzed using texture analysis. The analyzed data were statistically evaluated for every 20 pixels. These commands were deposited at the GitHub site as described in the Data Availability section. As a result, the profiles of saturation (colorfulness) and value (brightness) did not always differ among Gr1, Gr2, and Gr3 (Figs. S6 and S7). However, the profiles of Heu (color tone) for the original image were statistically different among Gr1, Gr2, and Gr3 (Fig. 6c) in the following pixel values: the pixel [2:22] (Gr1 vs Gr2, $p$=0.2624; and Gr1 vs Gr3, $p$=0.0114);

[12:32] (Gr1 vs Gr2, $p$=0.0354; and Gr1 vs Gr3, $p$=0.0008); [22:42] (Gr1 vs Gr2, $p$=0.0504; and Gr1 vs Gr3, $p$=0.0686); [106:126] (Gr1 vs Gr2, $p$=0.164876; and Gr1 vs Gr3, $p$=0.067624); [116:136] (Gr1 vs Gr2, $p$=0.041206; and Gr1 vs Gr3, $p$=0.012415); [126:146] (Gr1 vs Gr3, $p$=0.00183); [136:156] (Gr1 vs Gr3, $p$=0.00183); [146:166] (Gr1 vs Gr3, $p$=0.00183); and [156:176] (Gr1 vs Gr3, $p$=0.021453). Similar trends were also observed in Gaussian treated photographs (Fig. 6d) as follows: the pixels [2:22] (Gr1 vs Gr2, $p$=0.2016; and Gr1 vs Gr3, $p$=0.0097); [12:32] (Gr1 vs Gr2, $p$=0.0157; and Gr1 vs Gr3, $p$=0.0004); [22:42] (Gr1 vs Gr2, $p$=0.0483; and Gr1 vs Gr3, $p$=0.0721); [106:126] (Gr1 vs Gr2, $p$=0.141586; and Gr1 vs Gr3, $p$=0.054397); [116:136] (Gr1 vs Gr2, $p$=0.033731; and Gr1 vs Gr3, $p$=0.012415); [126:146] (Gr1 vs Gr3, $p$=0.00183); [136:156] (Gr1 vs Gr3, $p$=0.00183); [146:166] (Gr1 vs Gr3, $p$=0.00183); and [156:176] (Gr1 vs Gr3, $p$=0.021453) .

These results indicate that the qualitative differences in the surface of the fish in this study can be quantitatively evaluated using HSV analysis.

## Discussion

Today's environmental problems are a complex combination of factors, and in particular, it is necessary to use nitrogen and phosphorus efficiently and reduce the use of artificial antibiotics. Under such circumstances, the results of this research provide important information for the operation of high-value-added recycling technology that leads to the reduction of artificial antibiotics, and for the technology that can efficiently evaluate the added value using images of the surface of the fish body. Based on these observations, the potential mechanisms underlying the effects of thermophile-fermented compost were hypothesized as the conceptual image (Fig. 7). Specifically, the following two points were clarified: 1) administration of thermophilic compost reduces infections in fish; and 2) Such effects can be detected as quantitative differences not only by in-body evaluation but also by non-invasive evaluation using field photographs. Based on these multiple perspectives, this study is the first to assess the antipathogenic effects of thermophile-fermented compost using field photo diagnostic imaging.

The thermophile *Bacillaceae*-fermented compost used in this study appears to play an antipathogenic probiotic role in fish. In addition to the properties of general compost that promote the efficient use of nitrogen and phosphorus, thermophile-fermented compost is meant to possess a novel probiotic functionality as an alternative to antibiotics. In previous studies, the function of this compost was strongly influenced by the compost-derived *Caldibacillus hisashii*[42-44]. It cannot be ruled out that it may be an effect, but it is necessary to study what kind of fermentation thermophile affects fish physiological responses in the future.

Another point of this study is the noninvasive evaluation. In general, it is difficult to evaluate the body of a fish unless it is extremely sick in the field, and it is not possible to quantitatively and quickly evaluate it. However, the results of this study showed that a quantitative assessment can be easily performed using a single photograph via the developed commands.

Depending on the ratio calculation, there are various possibilities for the measurement. In addition, texture analysis can evaluate the degree of smoothness of images; therefore, it is less susceptible to the effects of light. In the field, this analytical method was considered useful because it was generally difficult to adjust the lighting in the photographs. Consequently, the tendency of entropy to increase under the experimental conditions of this study was confirmed. This trend was also confirmed in the edited image of the Gaussian model. It could be assumed that the conditions of the photograph were obscure.

Next, an HSV analysis was performed to rule out the effects of light. Under these experimental conditions, the effect of light was partly observed in V (Value, brightness) (Fig. S7). The results showed that there was a significant difference in V (Value, brightness), except for a part, but the effect was low, and a characteristically significant difference was observed in H (Heu, color tone) under these experimental conditions (Figs. 6c and 6d). These results indicated that the effect of composting can be statistically evaluated as the difference observed on the surface of the fish body. There is a possibility that changes that have been evaluated qualitatively thus far can be used as indicators by non-invasive analysis methods.

In the future, it is necessary to evaluate the relationship between various conditions, such as each fish species, growth stage, and environmental factors, but it is expected that it can be used as an index to quantitatively evaluate visual differences in the relevant experimental conditions. Furthermore, based on the results of this study, it means that the evaluation of compost, which can positively affect the circulation of nitrogen, phosphorus, and antibiotics, can be evaluated non-invasively, easily and quantitatively through changes in the surface of the fish body. In addition, aquaculture facilities-located areas are generally densely populated with several facilities, which could lead to the spread of infectious diseases. This area results in the excessive use of antibiotics, which has a negative impact on biodiversity. Therefore, it can be said that the model in this study is also an attempt to reduce the environmental load of antibiotics and simultaneously contribute to productivity and quality of fish by efficiently assessing the degree of smoothness and gloss of the body surface.

Thus, the approach of this study gives rise to a new perspective on ecosystem sustainability and a natural symbiotic society. Further research to evaluate the images efficiently could contribute not only in environmental studies, biology, and medicine, but also in other several fields.

## Acknowledgments

We are grateful to Seiichi Wada (Chuubushiryo Co. Ltd.), Toshihito Shinmyo (Keiyo Gas Energy Solution Co., Ltd.), and Jirou Matsumoto, Kazuo Ogawa (Sermas Co., Ltd.) for help with breeding facilities, sampling, and/or preparation of feed, to Teruno Nakaguma and Chitose Ishii (Sermas Co., Ltd.) for discussion on the imaging. Special thanks to Professor Hideaki Haneishi for providing technical advice on the imaging.

## Author contributions

Hirokuni Miyamoto and Hisashi Miyamoto conceived and designed the experiments; Shunsuke Ito performed the fish experiments; Hisashi Miyamoto and Motoaki Udagawa prepared the compost extract; Hirokuni Miyamoto, Shunsuke Ito, Takayuki Nagatsuka, and Takashi Satoh performed biological assay and the data arrangement; Hirokuni Miyamoto, Kenta Suzuki, Shingo Tamachi, and Shion Yamada performed the creation of commands for image analysis and the analysis; Hirokuni Miyamoto, Kenta Suzuki, Shingo Tamachi, Shion Yamada, and Jun Kikuchi analyzed the raw data.; Hirokuni Miyamoto, Kenta Suzuki, and Jun Kikuchi wrote the manuscript. All authors have read and approved the final manuscript.

## Data availability

Commands for image analysis have been registered on GitHub (https://github.com/hmiyamoto2000/program_tai3-main/tree/v0.1.1; https://github.com/hmiyamoto2000/program_tai1_com; https://github.com/hmiyamoto2000/program_tai1_Texture_main/tree/v1.0.6 ; https://github.com/hmiyamoto2000/program_tai1_HSV_main/tree/v1.0.5; and https://github.com/kecosz/cv2_jpg_preprocessing) and DOIs URL (10.5281/zenodo.15080985; 10.5281/zenodo.15081665; 10.5281/zenodo.15081663; 10.5281/zenodo.15081661; and 10.5281/zenodo.15081995). In addition, the photos used for the GitHub commands were stored on the figshare site and registered (10.6084/m9.figshare.28646039). All the data were stored in a data file (Data_file_fish.xlsx). If you need the information not listed here, please contact the corresponding authors.

## Funding

The Research and Development Support Program for Regional Revitalization of the Food Industry Sector of the Ministry of Agriculture, Forestry, and Fisheries of Japan (No. 22331219048)
Ministry of Economy, Trade, and Industry of Japan (No.21231219119)

## Competing interests

The authors declare no competing interests.

# Figure legends

## Fig. 1 Outline of this research

(a) Concept using recycled fermented feed. In this study, the anti-pathogenic effects of 0, 1%, and 5% of a 1/100 diluted solution of compost (compost extract) as a feed additive to basic feed are evaluated. (b) Plan for this study. A 70-day breeding test is conducted in advance; no compost extract is administered to the control group (Gr1) and 1% or 5% compost extract is administered to the test group (Gr2 and Gr3, respectively). After preliminary feeding, *Edwardsiella tarda* is injected intraperitoneally (0.3 ml of the solution with the pathogen, and the content was 4.3 x $10^6$ cfu/ml), the subsequent survival rate is evaluated. In addition, the physiological response and diagnostic imaging of the body surface are investigated for Gr1, Gr2, and Gr3 via *in vivo* internal evaluation and *in vivo* external evaluation. (c) Breeding results for the preparatory and subsequent periods with *Edwardsiella* exposure. Abbreviations are as follows: *i.p.*, intraperitoneal injection; Gr1 (Control), control group without compost extract; Gr2 (Extract 1%), the group administered 1% compost extract; Gr3 (Extract 5%), the group administered 5% compost extract; BW, body weight; FCR, feed conversion ratio (increased body weight per feed volume during the breeding period).

## Fig. 2 Evaluation of survival rate and physiological indicators after administration

(a) Survival curves obtained using the Kaplan-Meier method. Statistical analyses are performed using the Cox regression function to show the values of the log-rank test and likelihood ratio test in the graph. The table below shows the number of survivors on each of the five test days. (b) Serum complement activity before and after test initiation. The abbreviations are as follows: -70d, a beginning point in prior breeding period; 0d, just before the *Edwardsiella* injection; 28d, 28 days after the *Edwardsiella* injection. (c) Recovery time after anesthesia administration. 2-methylquinoline, an anesthetic that can be generally administered without the use of a syringe, is used.

## Fig. 3 Body length evaluation

(a) Workflow of analytical steps for body length measurements used in this study. (b) Photographs for each measurement step. The step (i) is performed to determine the measurement range. That is, the blue line, rgb (0, 0, 255), as the RGB values, connects X1 and X2 (standard body length, the center line in this study), as shown in Part I. Then, dots with different RGB values are placed at the base of the pectoral fins and at the front of the anus. The colors of these dots are set as follows: G1 in Part I, rgb (0, 255, 0); R1, 255:0:0 as the RGB value; and A1, rgb (254, 254, 0). Next, a dedicated command is used to draw lines perpendicular to the blue line, as shown in (ii). In Part II, the orthogonal lines G1, R1, and Z1 are shown. Thereafter, dots with different RGB values are manually placed at the intersection of the upper and lower surfaces of the fish body as shown in (iii). The colors of these dots are set as follows: Y1 and Y2 in Part II, rgb (255, 0, 255); Z2 in Part II, rgb (254, 254, 0); and X1 and X2, rgb (2, 0, 255). Finally, a developed command automatically calculates the ratio of body length to height in Part III of (b), as shown in (iv): R1, the ratio $(y_1\text{-}y_2)/(x_1\text{-}x_2)$; R2, the ratio $(z_1\text{-}z_2)/(x_1\text{-}x_2)$; and R3, the ratio$(z_1\text{-}z_2)/(y_1\text{-}y_2)$. (c) Evaluation of body length ratio for R1, R2, and R3. The asterisks indicated $p<0.05$.

## Fig. 4 Evaluation by texture analysis of body surface

(a) Workflow of the analytical steps for texture analysis used in this study. (b) Photograph of each texture measurement step. First, the positions of the mouth and tail are summarized in a CSV file based on the numerical values of ImageJ as the information to adjust the horizon level of the photo, as shown in (i). The background is changed to green as rgb (0,255,0), as seen in Part I. Next, a special command rotated the photo, and the fish bodies are parallel, as shown in Part II of step (ii). Then, in order to clarify the measurement range of the fish body as shown in step (iii), the above horizon adjustment is performed. The measuring range around the pectoral fins is automatically cut out as 1/3 of the area from the lip to the tail of the fish via special commands, as shown in Part III, and the texture analysis of this part is performed. Texture metrics include the contrast, dissimilarity, homogeneity, energy, correlation, and ASM and so forth. Furthermore, the analytical procedure is also used for Gaussian-treated photographs (gausian_gauss2.00 as a Graussion level). The conditions for the texture analysis and Gaussian treatment are described in the Materials and Methods section. (c) Texture feature metric candidates from the original and (d) those from the Gaussian-treated images. The candidates are selected via three types of machine learning (ML) algorithms.

## Fig. 5 Statistical comparison of texture features in normal and Gaussian photographs

(a) Main features of original photos. (b) Main features of the Gaussian-treated photos. In (a), the formula for the texture metrics (https://pyradiomics.readthedocs.io/en/latest/features.html#module-radiomics.glszm) is shown below each metric. The asterisks indicate the degree of significance: * $p<0.05$, ** $p<0.01$, and *** $p<0.001$.

## Fig. 6 HCV analysis of body surfaces

(a) Workflow of the analytical steps for HSV analysis used in this study. The process of splitting the photographs is performed in the same way as for texture analysis. A developed command then calculates the HSV values for the photo (marked as H, S, and V) in step (i). Next, the HSV profiles are statistically computed in the step (ii). A part of the profile is calculated for each of the 20 pixels. The abbreviations are as follows: H, Heu as color tone; S, Saturation as colorfulness; and V, Value as brightness. The Heu values from (b) the original photographs and (c) the Gaussian-treated photographs are shown for each group. The vertical Y axis represents frequency. The horizontal x-axis shows the range of 180 pixels, based on the OpenCV specification. The asterisks indicate the degree of significance: * $p<0.05$, ** $p<0.01$, and *** $p<0.001$.

## Fig. 7 Summary of this study

Compost fermented non-edible marine resources using thermophile *Bacillaceae* act as a disease-resistant feed additive as a highly valuable recycling technology. As the result of internal evaluation, growth promotion, mortality reduction, and the potential activities of serum complement and hepatic metabolism can be enhanced. Simultaneously, different degrees of smoothness and gloss of the fish body surface can be observed as the result of external evaluation: diagnostic imaging using field photographs statistically shows a significant increase in GLSZM ZoneEntropy as a metric for texture analysis and Heu value as a metric for HSV analysis.

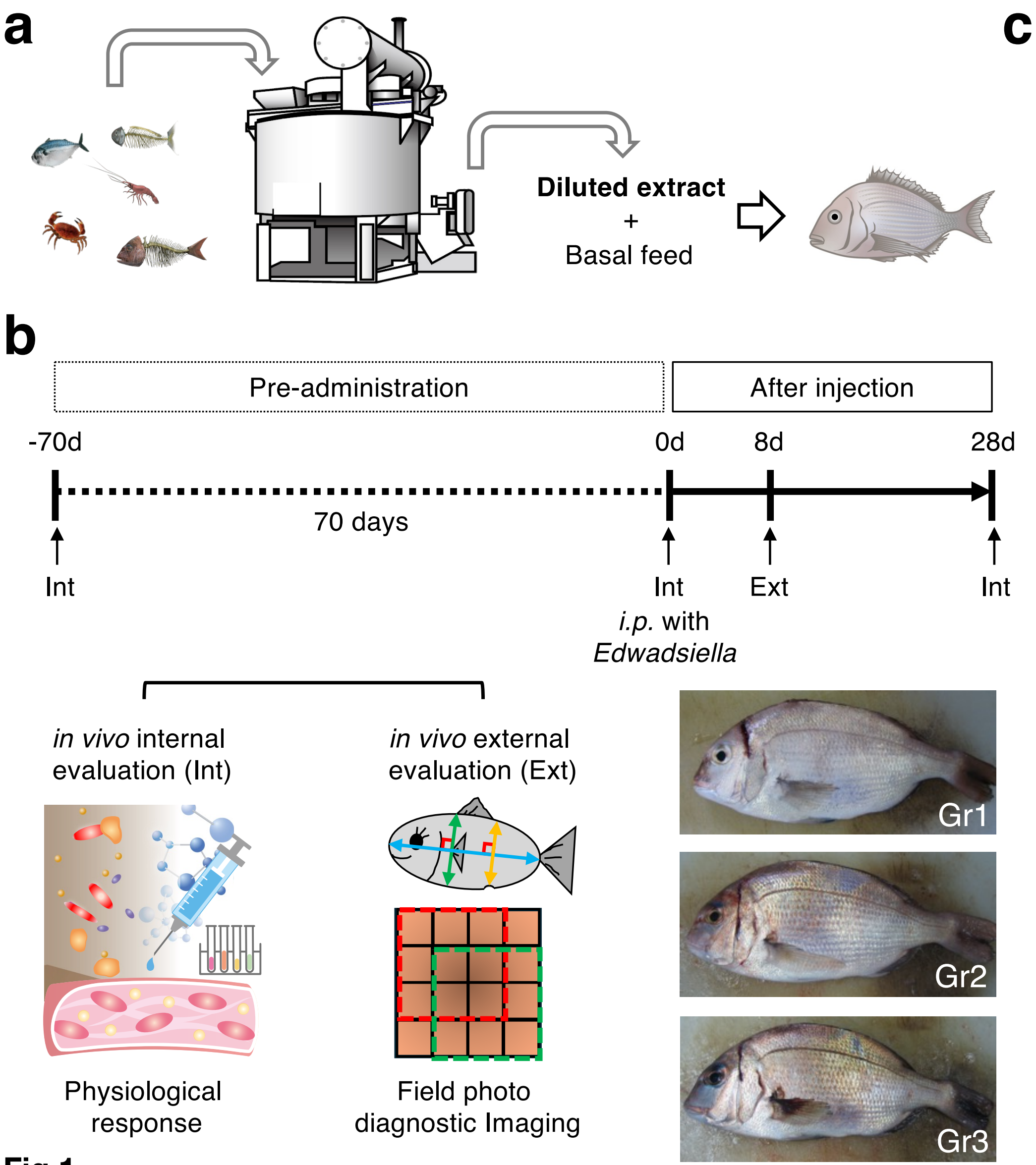

c

| Group name | Gr1 | Gr2 | Gr3 |
|---|---|---|---|
| Item | Physiological conditions | | |
| | Control | Extract 1% | Extract 5% |
| Before *E.taruda i.p.* | | | |
| Number of fish (-70d) | 15 | 15 | 15 |
| During of rearing (Day) | 70 | 70 | 70 |
| Finished total B.W. (g) | 3635 | 3620 | 3660 |
| Finished av. B.W. (g) | 242.3 | 241.3 | 244 |
| Increased B.W. (g) | 110.7 | 109.7 | 112.3 |
| Total feed weight (g) | 2439.2 | 2525.3 | 2553.7 |
| FCR | 0.681 | 0.651 | 0.660 |
| For Biophysical test | | | |
| Number of fish (0d) | 5 | 5 | 5 |
| Total B.W. (g) | 1007.5 | 1214.2 | 1046.5 |
| Averaged B.W. (g) | 201.5 | 242.84 | 209.3 |
| For *i.p.* injection test | | | |
| Number of fish (0d) | 10 | 10 | 10 |
| Total B.W. (g) (Beginning) | 2627.5 | 2405.8 | 2613.5 |
| Av. B.W. (g) (Beginning) | 262.8 | 240.6 | 261.4 |
| During of rearing (Day) | 28 | 28 | 28 |
| Number of survival fish (28d) | 4 | 9 | 8 |
| Finished total B.W. (g) | 936.8 | 2040.4 | 1956.5 |
| Finished av. B.W. (g) | 223.7 | 226.7 | 253.6 |

**Fig.1**

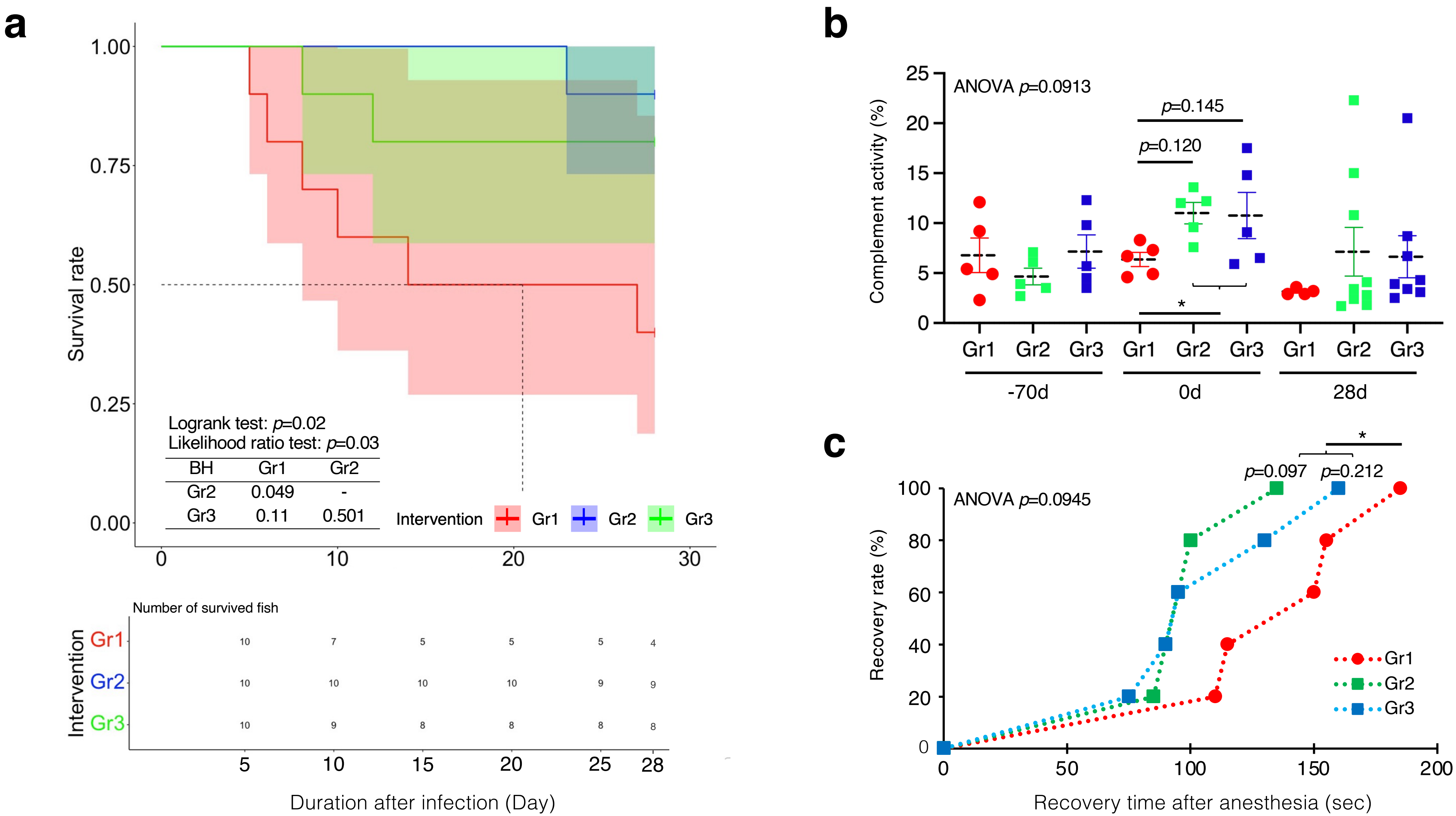

a
1.00
0.75
0.50
0.25
0.00
Survival rate
Logrank test: p=0.02
Likelihood ratio test: p=0.03
BH Gr1 Gr2
Gr2 0.049 -
Gr3 0.11 0.501
Intervention Gr1 Gr2 Gr3
0 10 20 30
Number of survived fish
Intervention
Gr1 10 7 5 5 5 4
Gr2 10 10 10 10 9 9
Gr3 10 9 8 8 8 8
5 10 15 20 25 28
Duration after infection (Day)
b
Complement activity (%)
25 20 15 10 5 0
ANOVA p=0.0913
p=0.145
p=0.120
*
Gr1 Gr2 Gr3 Gr1 Gr2 Gr3 Gr1 Gr2 Gr3
-70d
0d
28d
c
*
p=0.097 p=0.212
ANOVA p=0.0945
Recovery rate (%)
100 80 60 40 20 0
Gr1
Gr2
Gr3
0 50 100 150 200
Recovery time after anesthesia (sec)

**Fig.2**

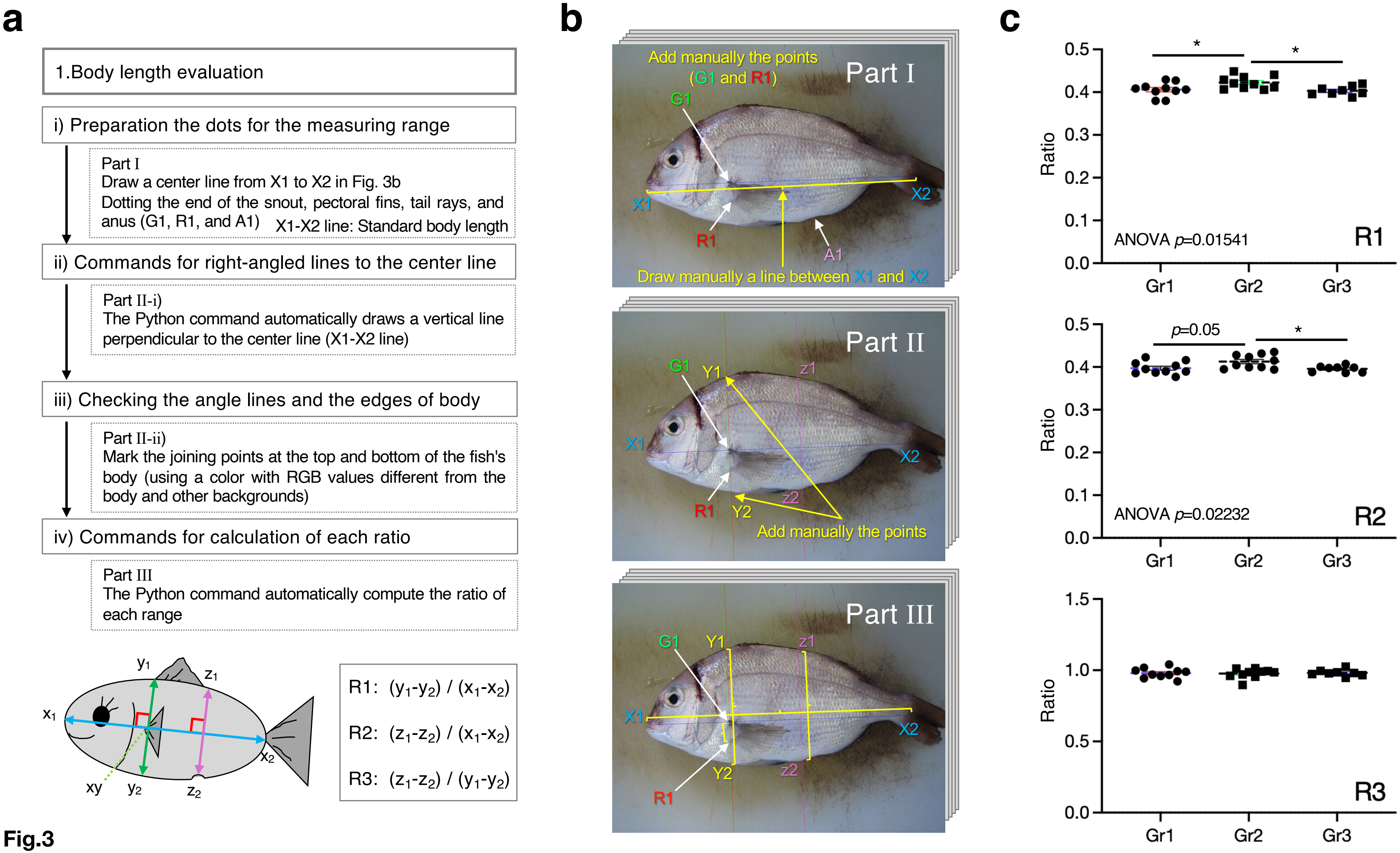
a
1.Body length evaluation
i) Preparation the dots for the measuring range
Part I
Draw a center line from X1 to X2 in Fig. 3b
Dotting the end of the snout, pectoral fins, tail rays, and anus (G1, R1, and A1)
X1-X2 line: Standard body length
ii) Commands for right-angled lines to the center line
Part II-i)
The Python command automatically draws a vertical line perpendicular to the center line (X1-X2 line)
iii) Checking the angle lines and the edges of body
Part II-ii)
Mark the joining points at the top and bottom of the fish's body (using a color with RGB values different from the body and other backgrounds)
iv) Commands for calculation of each ratio
Part III
The Python command automatically compute the ratio of each range
$y_1$
$z_1$
$x_1$
$x_2$
xy
$y_2$
$z_2$
R1: $(y_1-y_2) / (x_1-x_2)$
R2: $(z_1-z_2) / (x_1-x_2)$
R3: $(z_1-z_2) / (y_1-y_2)$
b
Add manually the points (G1 and R1)
Part I
G1
X1
X2
R1
A1
Draw manually a line between X1 and X2
Part II
G1
Y1
z1
X1
X2
R1
Y2
z2
Add manually the points
Part III
G1
Y1
z1
X1
X2
Y2
z2
R1
c
0.5
0.4
0.3
0.2
0.1
0.0
Ratio
*
*
ANOVA $p$=0.01541
R1
Gr1
Gr2
Gr3
0.5
0.4
0.3
0.2
0.1
0.0
Ratio
$p$=0.05
*
ANOVA $p$=0.02232
R2
Gr1
Gr2
Gr3
1.5
1.0
0.5
0.0
Ratio
R3
Gr1
Gr2
Gr3

**Fig.3**

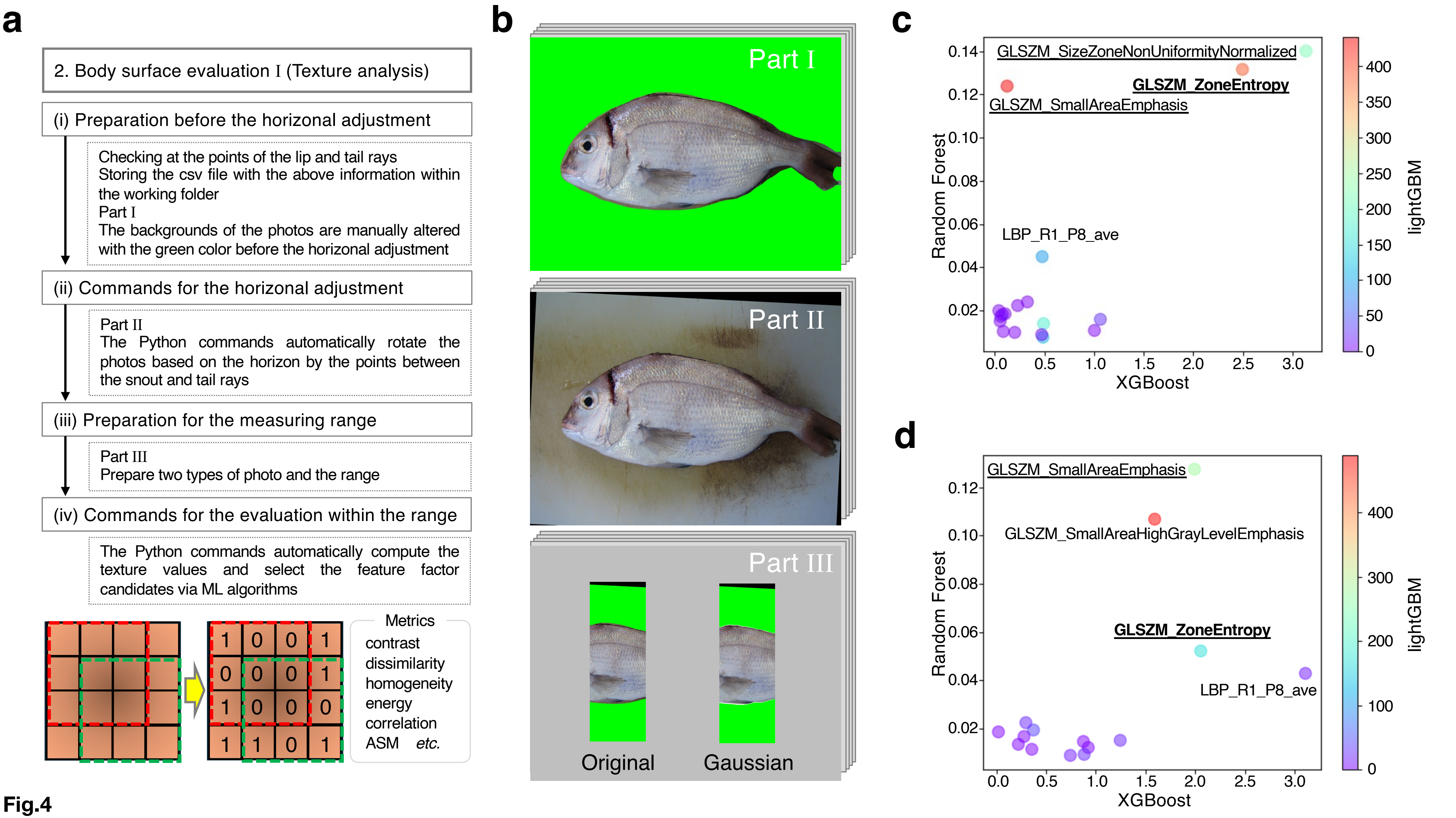

a
2. Body surface evaluation I (Texture analysis)
(i) Preparation before the horizonal adjustment
Checking at the points of the lip and tail rays
Storing the csv file with the above information within the working folder
Part I
The backgrounds of the photos are manually altered with the green color before the horizonal adjustment
(ii) Commands for the horizonal adjustment
Part II
The Python commands automatically rotate the photos based on the horizon by the points between the snout and tail rays
(iii) Preparation for the measuring range
Part III
Prepare two types of photo and the range
(iv) Commands for the evaluation within the range
The Python commands automatically compute the texture values and select the feature factor candidates via ML algorithms
1 0 0 1
0 0 0 1
1 0 0 0
1 1 0 1
Metrics
contrast
dissimilarity
homogeneity
energy
correlation
ASM etc.
b
Part I
Part II
Part III
Original
Gaussian
c
GLSZM_SizeZoneNonUniformityNormalized
GLSZM_ZoneEntropy
GLSZM_SmallAreaEmphasis
LBP_R1_P8_ave
Random Forest
XGBoost
lightGBM
0.14
0.12
0.10
0.08
0.06
0.04
0.02
0.0
0.5
1.0
1.5
2.0
2.5
3.0
400
350
300
250
200
150
100
50
0
d
GLSZM_SmallAreaEmphasis
GLSZM_SmallAreaHighGrayLevelEmphasis
GLSZM_ZoneEntropy
LBP_R1_P8_ave
Random Forest
XGBoost
lightGBM
0.12
0.10
0.08
0.06
0.04
0.02
0.0
0.5
1.0
1.5
2.0
2.5
3.0
400
300
200
100
0

Fig.4

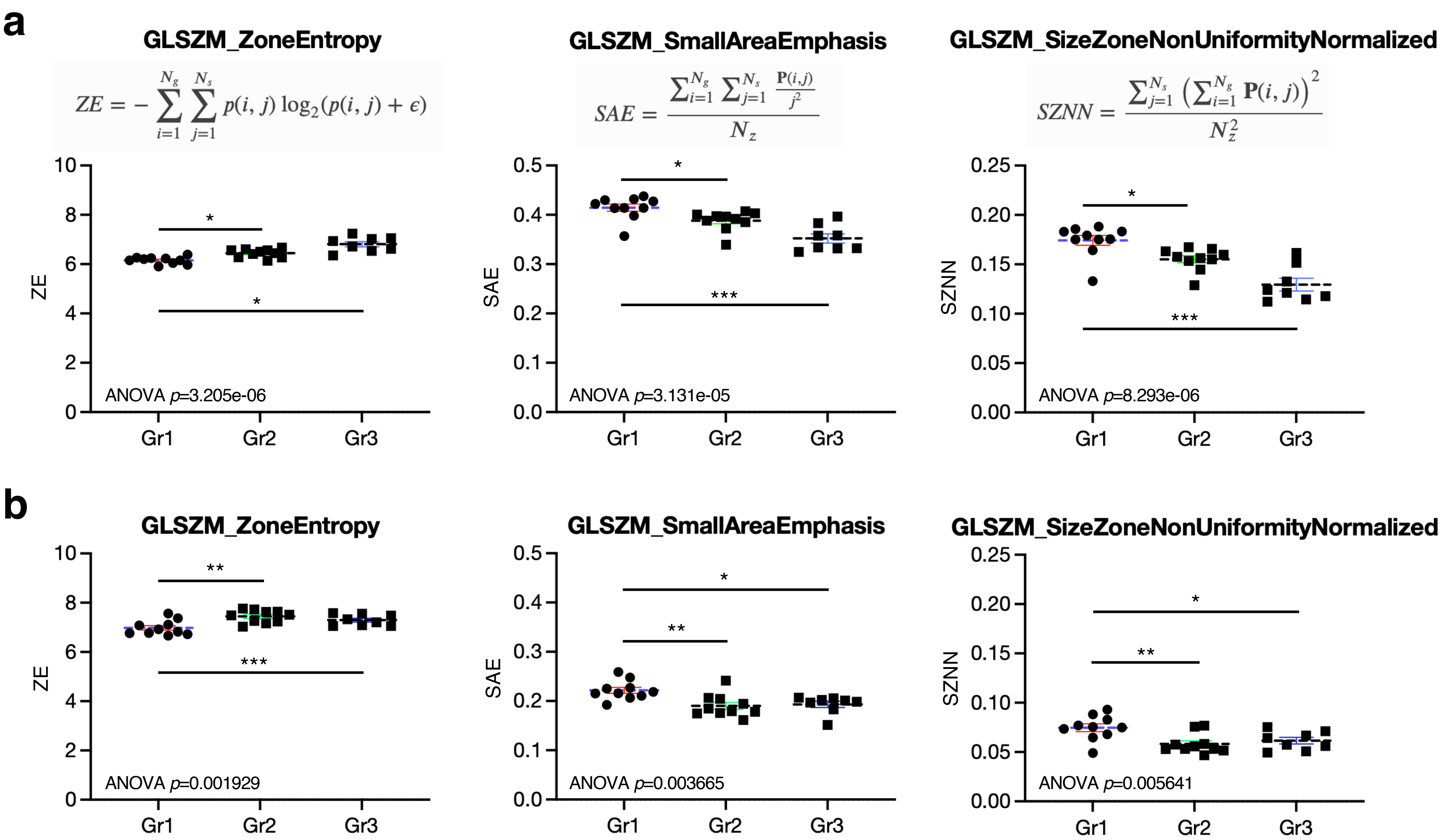
a
GLSZM_ZoneEntropy
$ZE = -\sum_{i=1}^{N_g}\sum_{j=1}^{N_s} p(i,j)\log_2(p(i,j)+\epsilon)$
ZE
ANOVA p=3.205e-06
Gr1
Gr2
Gr3
GLSZM_SmallAreaEmphasis
$SAE = \frac{\sum_{i=1}^{N_g}\sum_{j=1}^{N_s}\frac{\mathbf{P}(i,j)}{j^2}}{N_z}$
SAE
ANOVA p=3.131e-05
GLSZM_SizeZoneNonUniformityNormalized
$SZNN = \frac{\sum_{j=1}^{N_s}\left(\sum_{i=1}^{N_g}\mathbf{P}(i,j)\right)^2}{N_z^2}$
SZNN
ANOVA p=8.293e-06
b
GLSZM_ZoneEntropy
ANOVA p=0.001929
GLSZM_SmallAreaEmphasis
ANOVA p=0.003665
GLSZM_SizeZoneNonUniformityNormalized
ANOVA p=0.005641

**Fig.5**

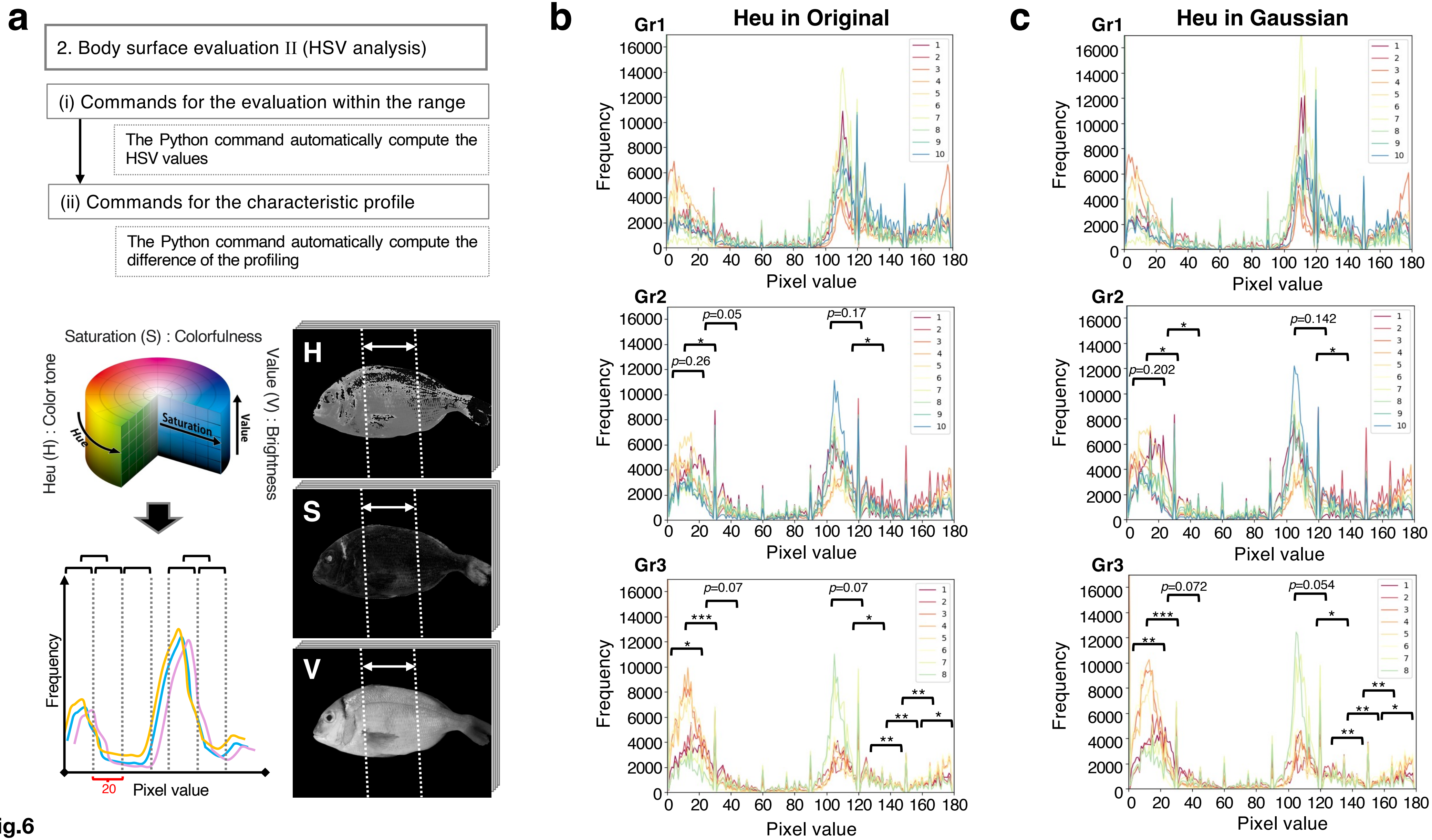
a
2. Body surface evaluation II (HSV analysis)
(i) Commands for the evaluation within the range
The Python command automatically compute the HSV values
(ii) Commands for the characteristic profile
The Python command automatically compute the difference of the profiling
Saturation (S) : Colorfulness
Heu (H) : Color tone
Value (V) : Brightness
Hue
Saturation
Value
H
S
V
Frequency
20
Pixel value
b
Heu in Original
Gr1
Gr2
Gr3
Frequency
Pixel value
p=0.05
p=0.26
p=0.17
p=0.07
c
Heu in Gaussian
p=0.202
p=0.142
p=0.072
p=0.054

Fig.6

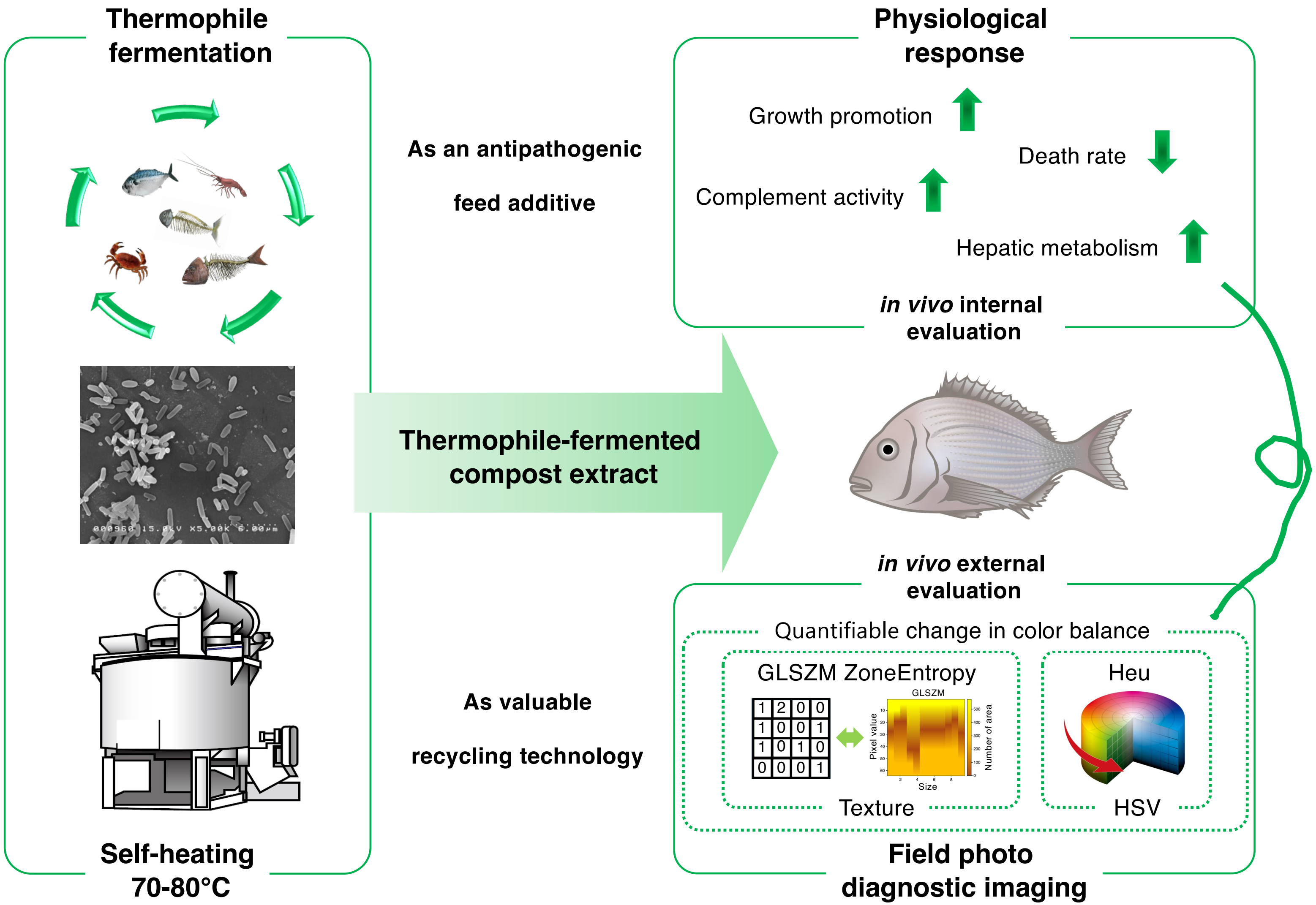
Thermophile fermentation
Self-heating 70-80°C
As an antipathogenic feed additive
As valuable recycling technology
Thermophile-fermented compost extract
Physiological response
Growth promotion
Death rate
Complement activity
Hepatic metabolism
in vivo internal evaluation
in vivo external evaluation
Quantifiable change in color balance
GLSZM ZoneEntropy
Texture
Heu
HSV
Field photo diagnostic imaging

Fig.7

# Supplementary Information

## A putative model for sustainable fisheries driven by non-invasive diagnostic imaging

Hirokuni Miyamoto[1,2,3,4,5*], Shunsuke Ito[6], Kenta Suzuki[7,8], Shingo Tamachi[9], Shion Yamada[10], Takayuki Nagatsuka[3,4], Takashi Satoh[11], Motoaki Udagawa[12], Hisashi Miyamoto[13], Hiroshi Ohno[2], Jun Kikuchi[5,14*]

*1. Graduate School of Horticulture, Chiba University, Matsudo, Chiba 271-8501, Japan*
*2. RIKEN Center for Integrative Medical Sciences, Yokohama, Kanagawa 230-0045, Japan*
*3. Japan Eco-science (Nikkan Kagaku) Co. Ltd., Chiba, Chiba 263-8522, Japan*
*4. Sermas Co., Ltd., Ichikawa, Chiba 272-0033, Japan*
*5. Graduate School of Medical Life Science, Yokohama City University, Yokohama, Kanagawa 230-0045, Japan*
*6. Chuubushiryo Co. Ltd., Oobu, Aichi 272-0033, Japan*
*7. RIKEN, BioResource Research Center, Tsukuba, Ibaraki 305-0074, Japan*
*8. Institute for Multidisciplinary Sciences, Yokohama National University, Yokohama, Kanagawa 240-8501, Japan*
*9. Center for Frontier Medical Engineering, Chiba University, Chiba, Chiba 263-8522,Japan*
*10. Faculty of Engineering, Chiba University, Chiba, Chiba 263-8522,Japan*
*11. Division of Hematology, Kitasato University School of Allied Health Sciences, Sagamihara, Kanagawa 252-0373, Japan*
*12. Keiyo Gas Energy Solution Co. Ltd., Ichikawa, Chiba 272-0033, Japan*
*13. Miroku Co. Ltd., Kitsuki, Oita 873-0021, Japan*
*14. RIKEN Center for Sustainable Resource Science, Yokohama, Kanagawa 230-0045, Japan*

*: Cocorrespondence

# Contents

**a**

| Group | Gr1 | Gr2 | Gr3 |
|---|---|---|---|
| Number of fish at the start | 15 | 15 | 15 |
| Total B.W. of fish at the start (g) | 1975 | 1960 | 1970 |
| Average B.W. of fish at the start (g) | 132 | 131 | 131 |
| **1st** | | | |
| During of rearing (days) | 14 | 14 | 14 |
| Finished total B.W. (g) | 2245 | 2285 | 2345 |
| Finished average B.W. (g) | 150 | 152 | 156 |
| Increased B.W. (g) | 270 | 325 | 375 |
| Feed weight (g) | 422 | 442 | 480 |
| FCR | 0.641 | 0.735 | 0.781 |
| **2nd** | | | |
| During of rearing (days) | 14 | 14 | 14 |
| Finished total B.W. (g) | 2485 | 2590 | 2655 |
| Finished average B.W. (g) | 166 | 173 | 177 |
| Increased B.W. (g) | 240 | 305 | 310 |
| Feed weight (g) | 407 | 484 | 476 |
| FCR | 0.589 | 0.630 | 0.652 |
| **3rd** | | | |
| During of rearing (days) | 14 | 14 | 14 |
| Finished total B.W. (g) | 2895 | 2895 | 2990 |
| Finished average B.W. (g) | 193 | 193 | 199 |
| Increased B.W. (g) | 410 | 305 | 335 |
| Feed weight (g) | 498 | 457 | 493 |
| FCR | 0.824 | 0.668 | 0.679 |
| 4th | | | |
| During of rearing (days) | 14 | 14 | 14 |
| Finished total B.W. (g) | 3330 | 3355 | 3405 |
| Finished average B.W. (g) | 222 | 224 | 227 |
| Increased B.W. (g) | 435 | 460 | 415 |
| Feed weight (g) | 696 | 685 | 730 |
| FCR | 0.625 | 0.672 | 0.568 |
| **5th** | | | |
| During of rearing (days) | 14 | 14 | 14 |
| Finished total B.W. (g) | 3635 | 3620 | 3660 |
| Finished average B.W. (g) | 242 | 241 | 244 |
| Increased B.W. (g) | 305 | 265 | 255 |
| Feed weight (g) | 417 | 457 | 374 |
| FCR | 0.732 | 0.579 | 0.681 |

**b**

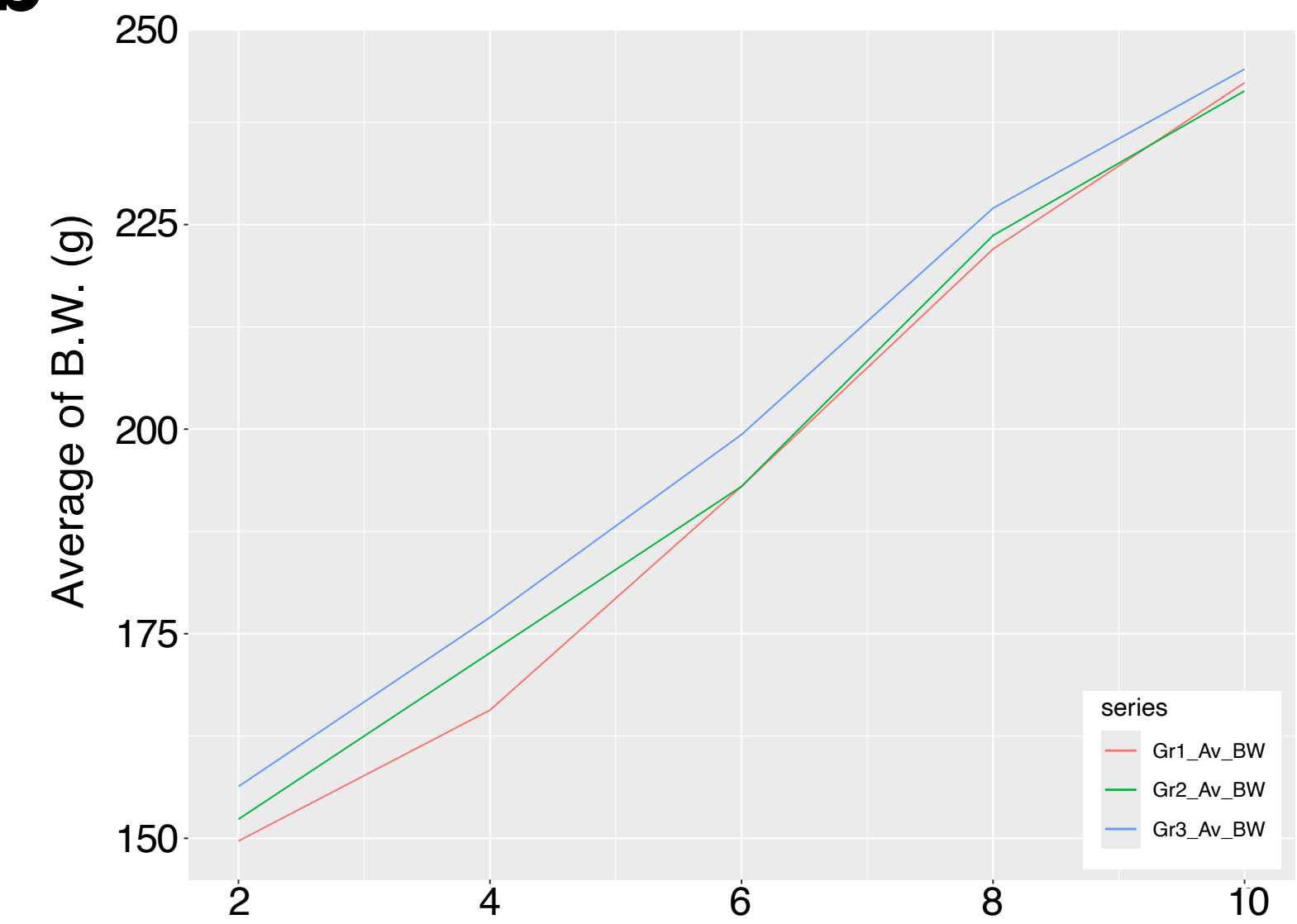

**c**

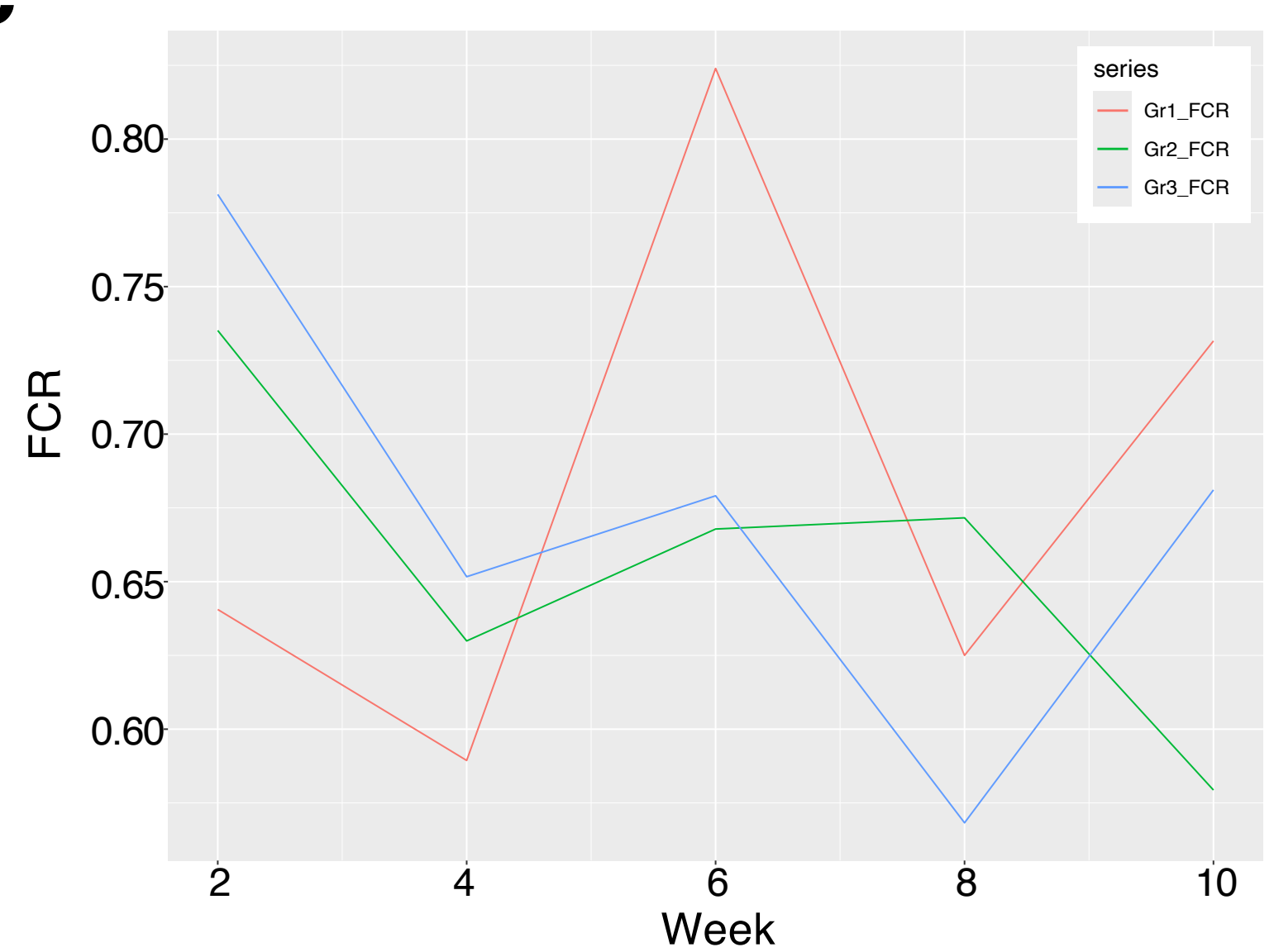

**Fig. S1**

**Body conditions of seabreams before *Edwadsiella* exposure.**

Physiological data of seabreams in the following treatment groups before intraperitoneal injection with pathogenic *Edwadsiella*: Gr1, control group (normal feeding conditions); Gr2, the group administered 1% compost extract; and Gr3, the group administered 5% compost extract. Panel (a) shows the table describing every 14 days for 70 days before the injection (1$^{st}$–5$^{th}$ as five growth stages). Panels (b) and (c) show the average B. W. and FCR profiles of fish, respectively. The abbreviations are as follows: Finished total B.W., sum of body weight of all fish in the group until the targeted stage; Finished average B. W., average body weight per fish in the group until the targeted stage; increased B.W., body weight at each growth stage; Feed weight, feed weight at each growth stage; and FCR, feed conversion ratio at each growth stage.

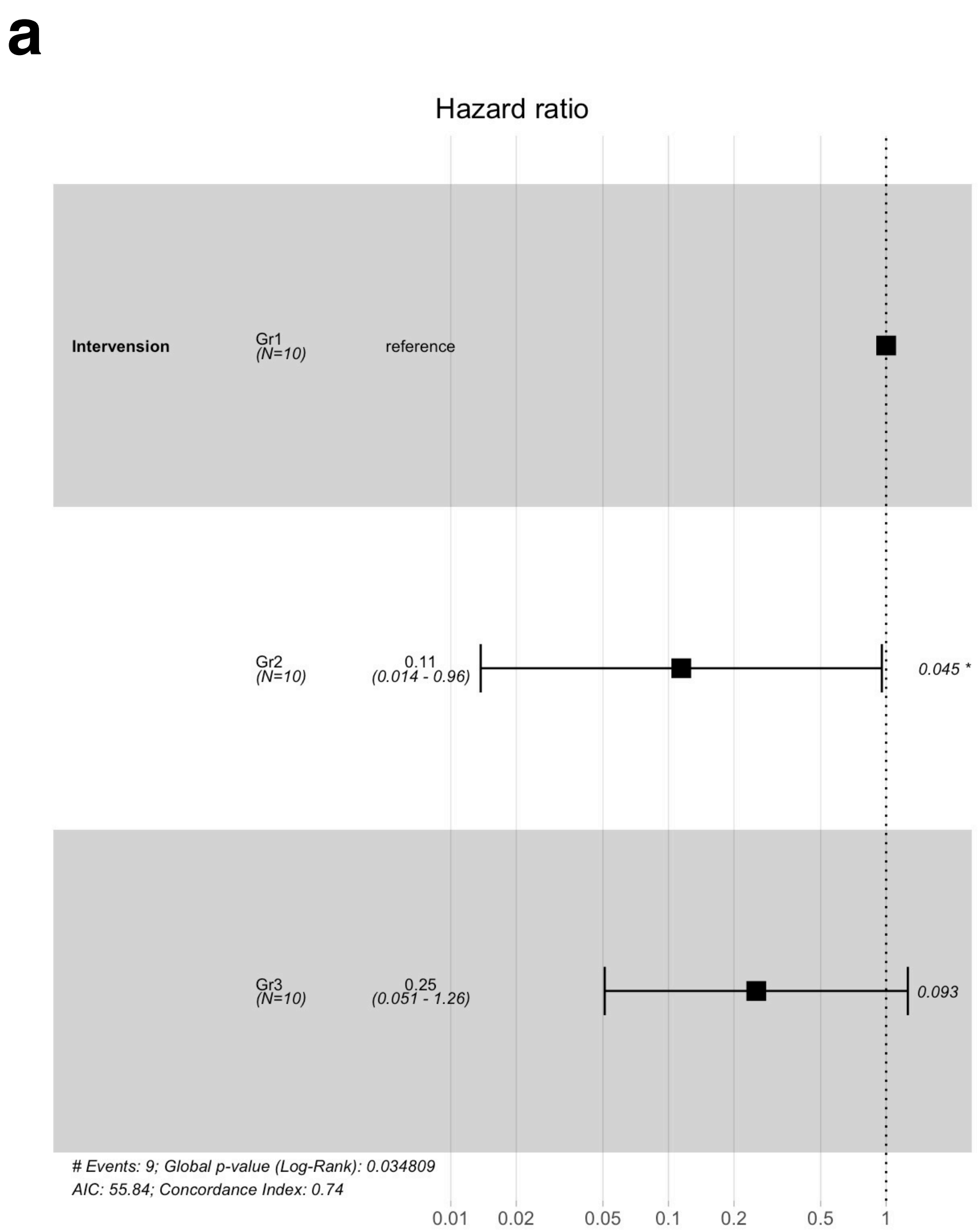

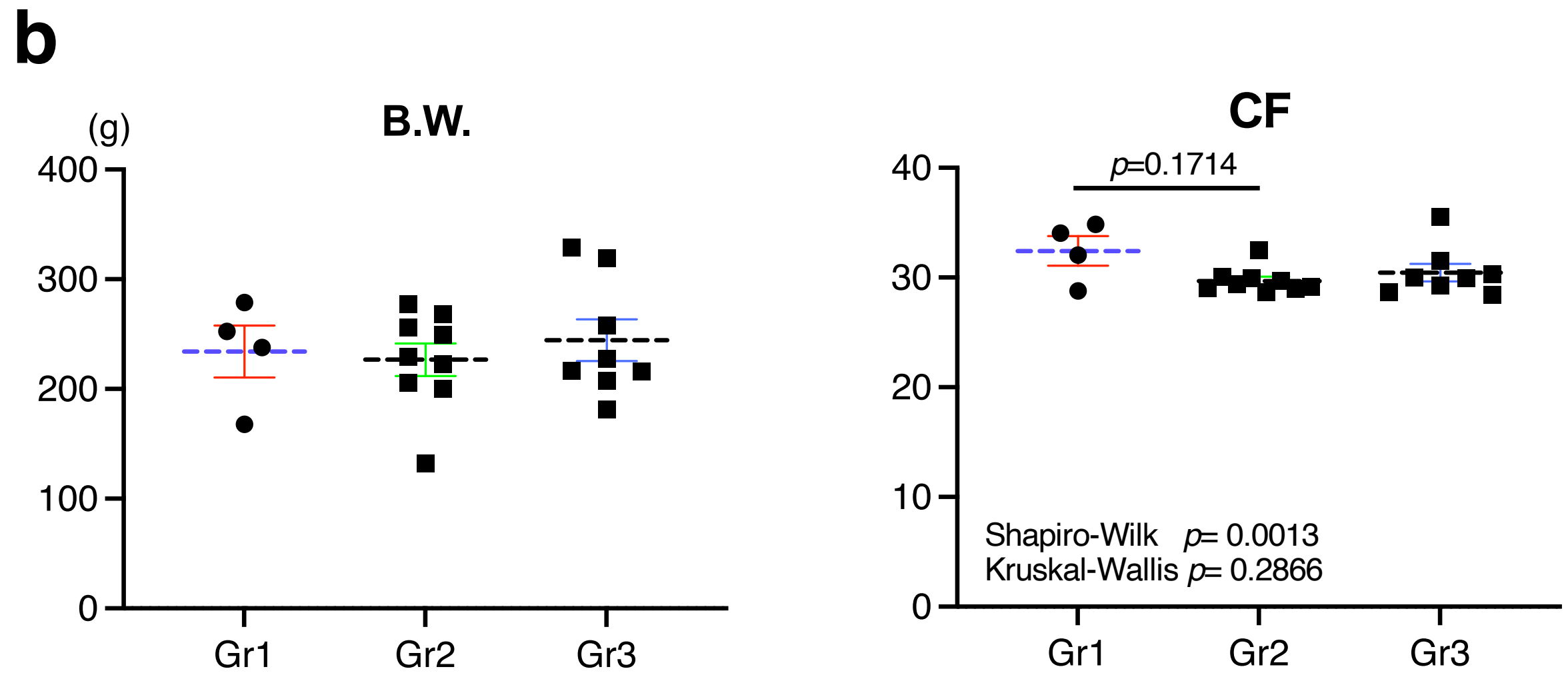

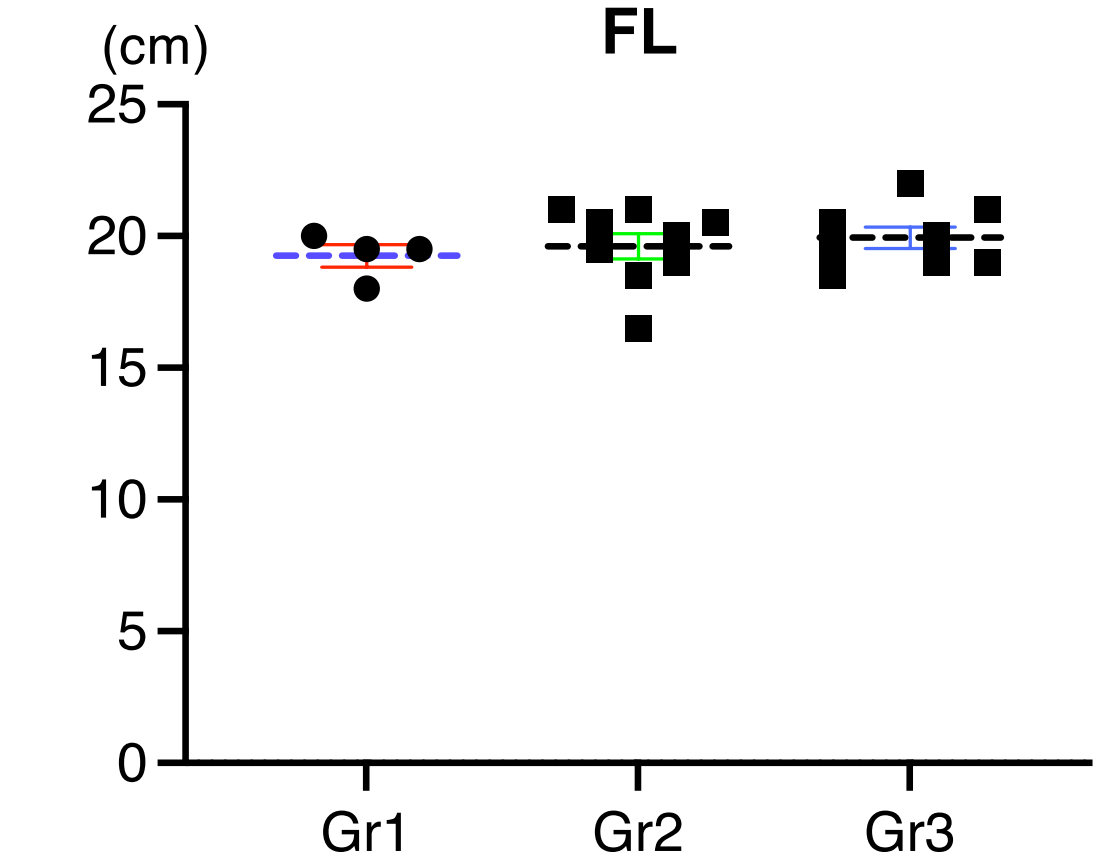

**Fig. S2**

**Physiological conditions of seabreams after *Edwadsiella* exposure.**

(a) Forest plot showing hazard ratio of three groups for the survival curve in Fig. 2a. The plot is visualized based on the cox regression model. (b) Body data of seabream following intraperitoneal injection of pathogenic *Edwadsiella*. The abbreviations are as follows: Gr1, control group (normal feeding conditions); Gr2, the group administered 1% compost extract; and Gr3, the group administered 5% compost extract; B.W., body weight of fish in the group; FL, Fork length; and CF (condition factor), the values calculated via (1000 × BW) / (FL × FL × FL).

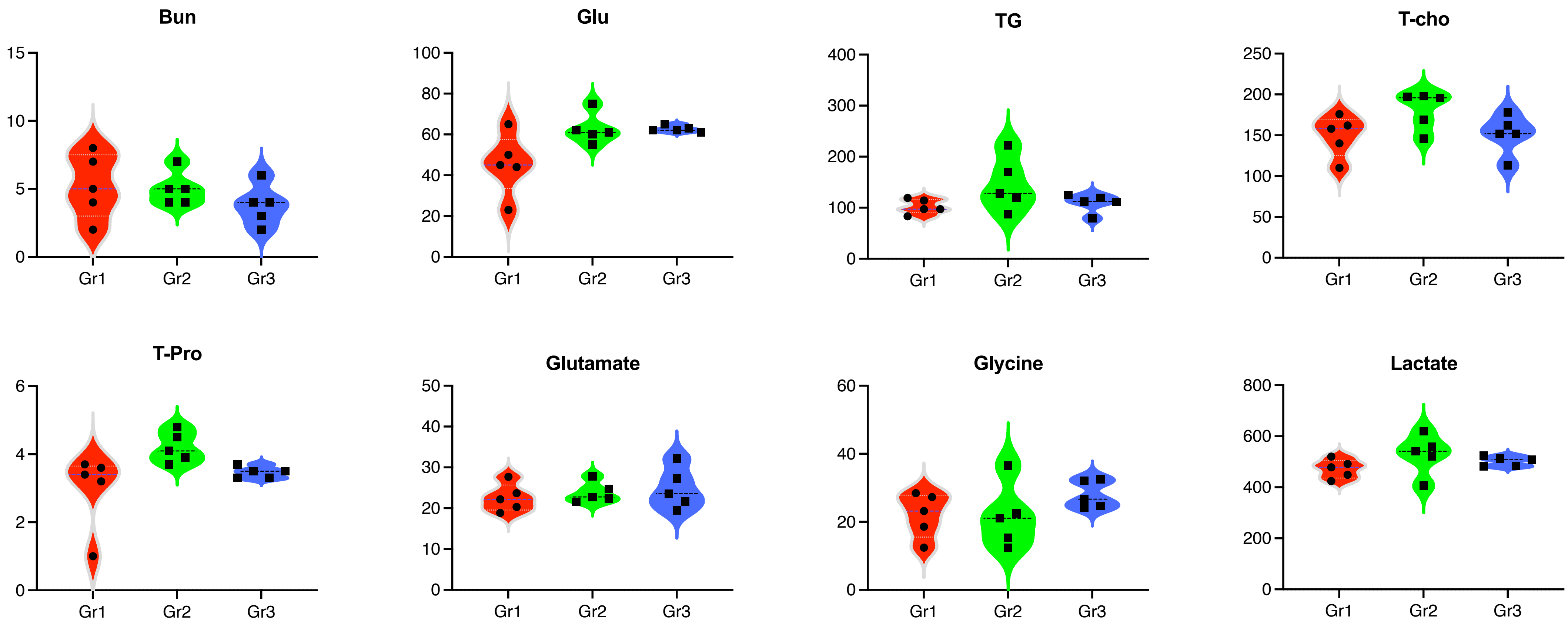

**Fig. S3**

**Physiochemical conditions of seabreams before exposure to Edwadsiella**

Physiological data of sea bream in the following treatment groups before intraperitoneal injection with pathogenic *Edwadsiella*: Gr1, control group (normal feeding conditions); Gr2, the group administered 1% compost extract; and Gr3, the group administered 5% compost extract. The abbreviations are as follows: Bun, serum uric nitrate; Glu, serum glucose; TG, serum triacylglyceride; T-cho, serum total cholesterol; T-Pro, serum total protein; Glutamate, glutamate contents in the muscle; Glycine, glycine contents in the muscle; lactate, lactate contents in the muscle.

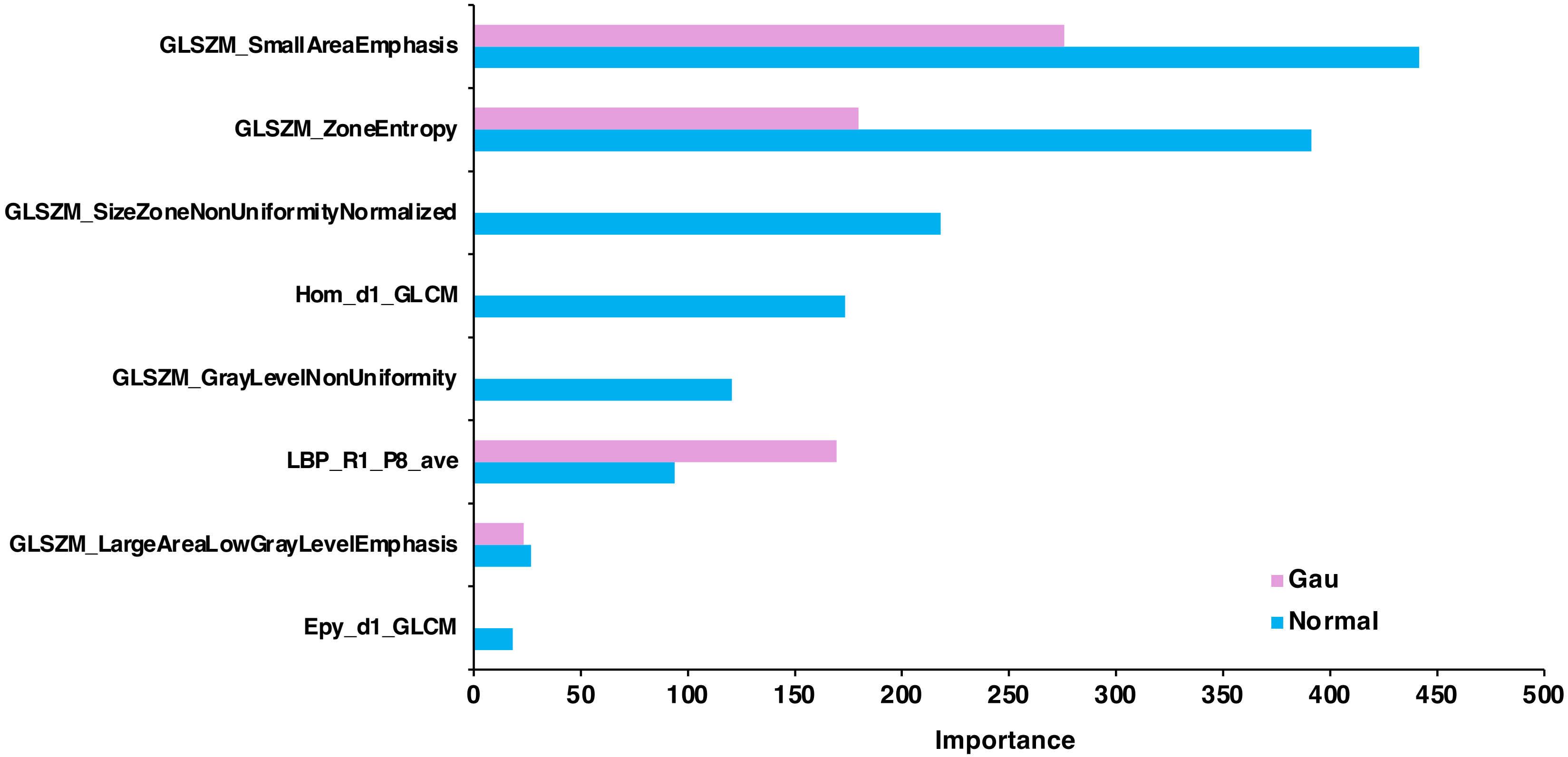

**Fig. S4**

**Comparison of LightGBM-selected features of texture indices on the surface of seabreams**

The texture indices of the photograph and the Gaussian-treated photo are compared with the features selected by LightGBM. Normal (light blue square) and Gau (purple square) indicate the degree of features from the normal (original) and Gaussian photos, respectively.

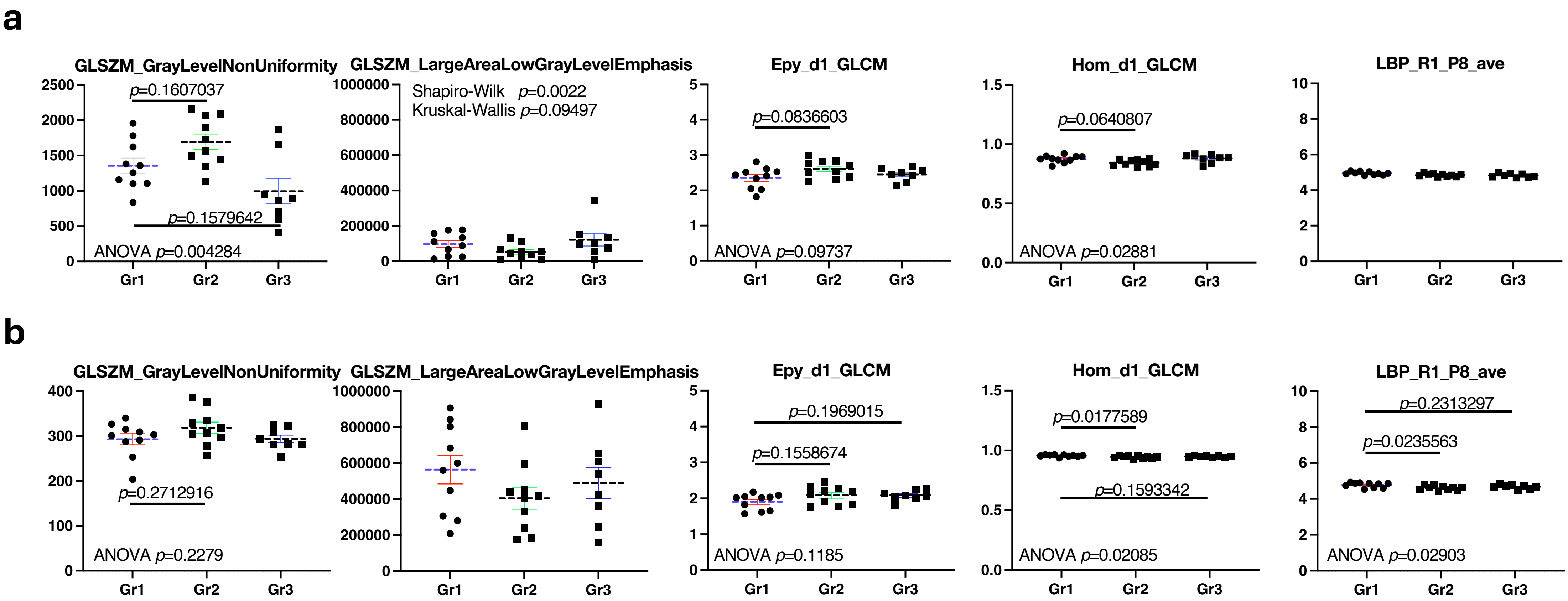

**Fig. S5**

**Values of the LightGBM-feature selected texture indices on the surface of the seabreams.**

The texture indices of (a) the photographs themselves and (b) the Gaussian processed photographs are evaluated by three types of machine learning algorithms, and the features selected by LightGBM are statistically compared. The unit of Y axis shows the degree of each metric.

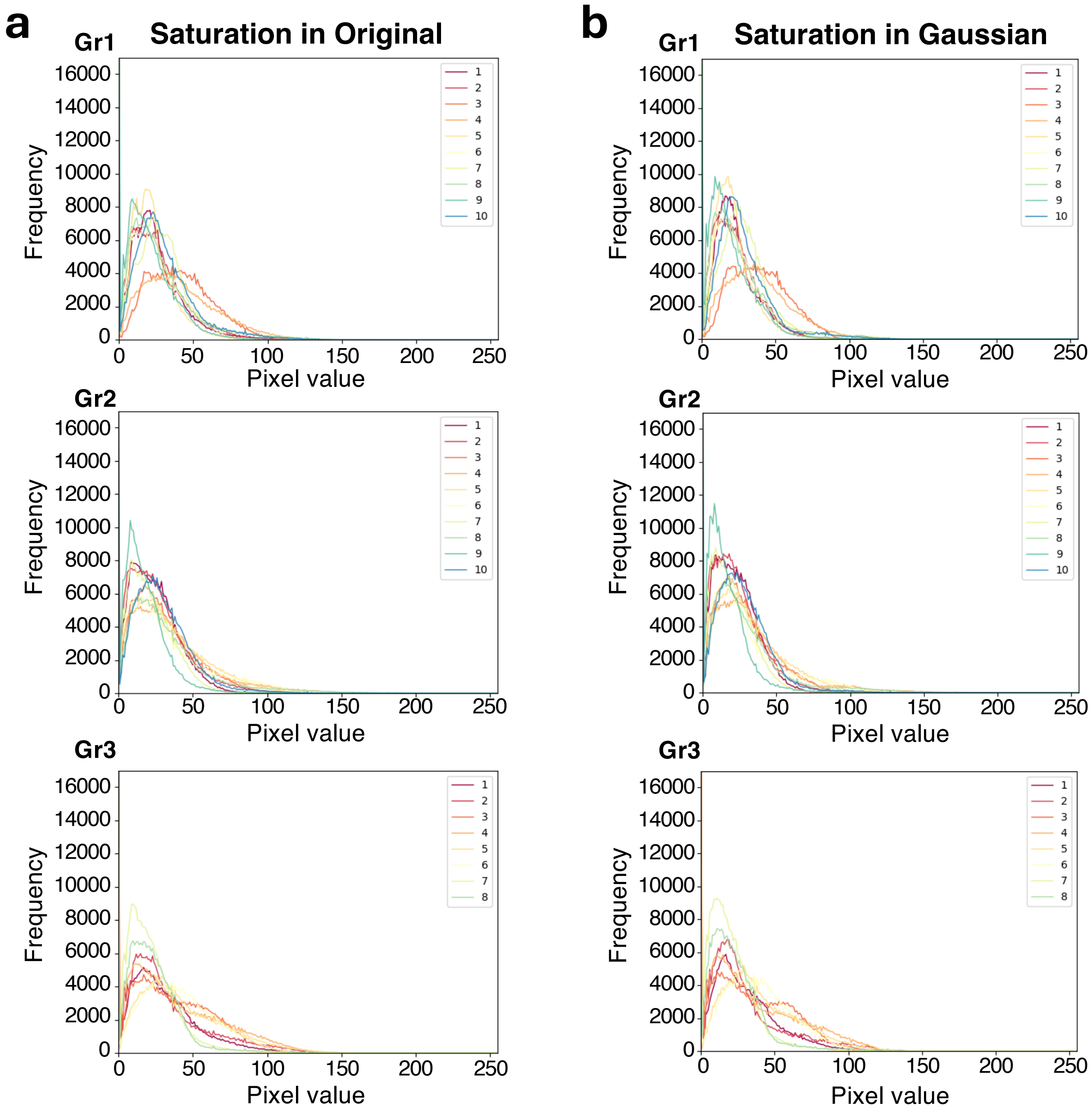

**Fig. S6**

**The values of saturation on the surface of seabreams via HSV analysis**

The saturation values from (a) the original photographs and (b) the Gaussian-treated photographs are shown for each group. The vertical Y axis represents frequency. The horizontal x-axis shows a range of 255 pixels, based on the OpenCV specification. All statistical analyses have $p>0.05$.

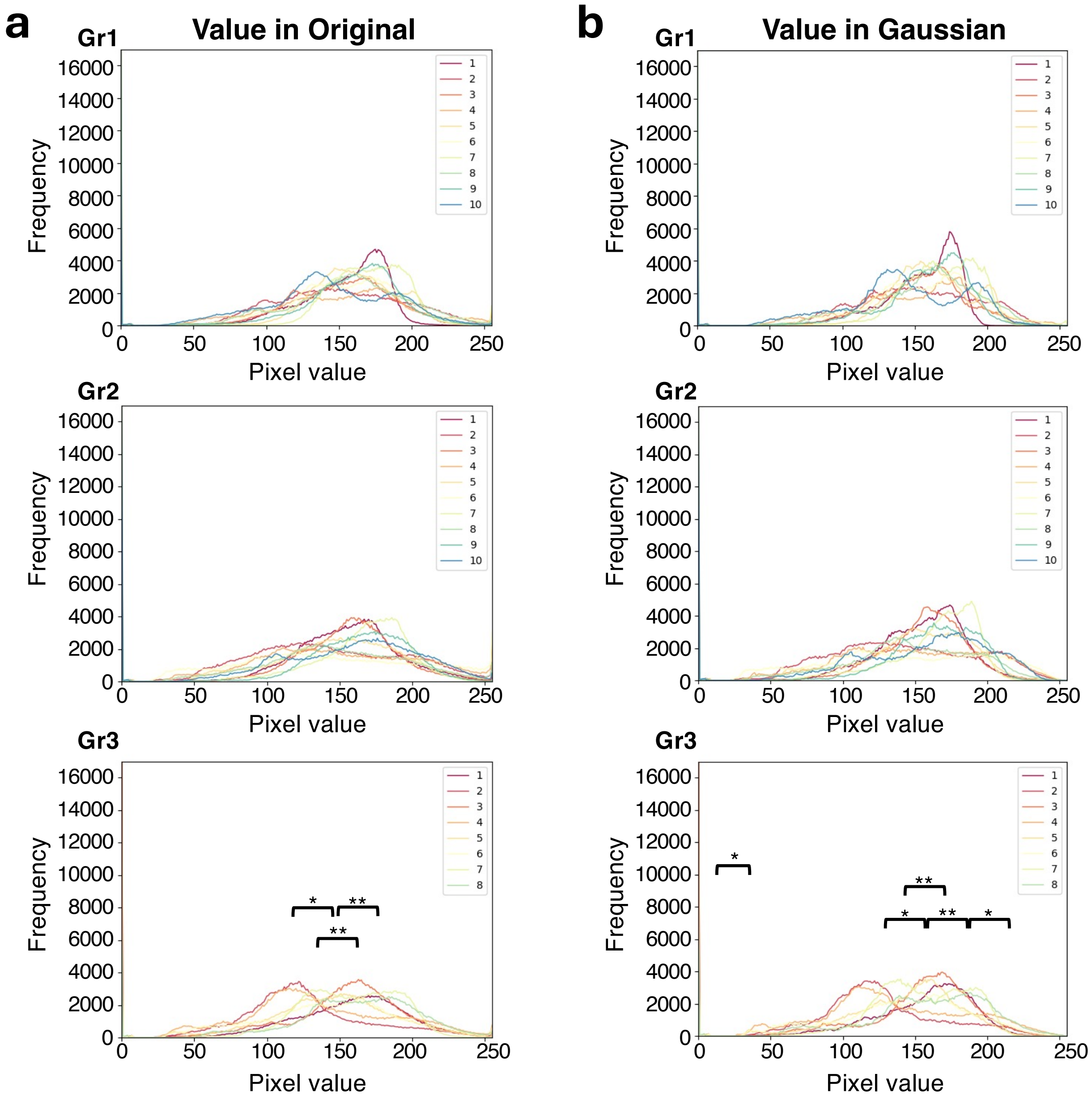

**Fig. S7**

**The values of brightness on the surface of seabreams via HSV analysis**

Brightness values from (a) the original photos and (b) the Gaussian-treated photos are shown for each group. The vertical Y axis represents frequency. The horizontal x-axis shows a range of 255 pixels, based on the OpenCV specification. Asterisks indicate the degree of significance: *, $p<0.05$.